\documentclass{aa}
\usepackage{graphicx}
\usepackage{bm}
\usepackage{times}
\usepackage{amsmath}
\usepackage{amssymb}
\usepackage{natbib}
\usepackage{color}
\bibpunct{(}{)}{;}{a}{}{,}
\graphicspath{{./fig/}{./png/}}

\newcommand{\EQ}{\begin{equation}}
\newcommand{\EE}{\end{equation}}
\newcommand{\EQA}{\begin{eqnarray}}
\newcommand{\EEA}{\end{eqnarray}}
\newcommand{\brac}[1]{\langle #1 \rangle}
\newcommand{\pd}{\partial}

\newcommand{\ve}[1]{\boldsymbol{#1}}
\newcommand{\mean}[1]{\overline{#1}}
\newcommand{\meanv}[1]{\overline{\bm #1}}

\newcommand{\nut}{\nu_{\rm t}}

\newcommand{\urms}{u_{\rm rms}}

\newcommand{\kef}{k_{\rm f}}

\newcommand{\chiSGS}{\chi_{\rm SGS}}
\newcommand{\chiSGSm}{\mean\chi_{\rm SGS}}

\newcommand{\Ta}{{\rm Ta}}

\newcommand{\Pra}{{\rm Pr}}
\newcommand{\PraSGS}{{\rm Pr}_{\rm SGS}}
\newcommand{\Ra}{{\rm Ra}}

\newcommand{\Rey}{{\rm Re}}
\newcommand{\Co}{{\rm Co}}

\newcommand{\Roc}{{\rm Ro}_{\rm c}}

\newcommand{\qij}{Q_{ij}}

\newcommand{\qrp}{Q_{r\phi}}
\newcommand{\qtp}{Q_{\theta\phi}}

\newcommand{\LamV}{\Lambda_{\rm V}}
\newcommand{\LamH}{\Lambda_{\rm H}}

{}
\def\onethird{{\textstyle{1\over3}}}
\def\onehalf{{\textstyle{1\over2}}}

\newcommand{\kg}{\,{\rm kg}}
\newcommand{\K}{\,{\rm K}}
\newcommand{\s}{\,{\rm s}}
\newcommand{\m}{\,{\rm m}}

\begin{document}

\authorrunning{K\"apyl\"a et al.}
\titlerunning{Bistable stellar differential rotation}

   \title{Confirmation of bistable stellar differential rotation profiles}

   \author{P. J. K\"apyl\"a
          \inst{1,2,3}
          \and
          M. J. K\"apyl\"a
          \inst{2}
           \and 
          A. Brandenburg
          \inst{3,4}
          }

   \offprints{\email{petri.kapyla@helsinki.fi}
              }

   \institute{Department of Physics, Gustaf H\"allstr\"omin katu 2a 
              (PO Box 64), FI-00014 University of Helsinki, Finland
         \and ReSoLVE Centre of Excellence, Department of Information and Computer Science, Aalto University, PO Box 15400, FI-00076 Aalto, Finland
         \and NORDITA, KTH Royal Institute of Technology and Stockholm University, Roslagstullsbacken 23, SE-10691 Stockholm, Sweden
         \and Department of Astronomy, Stockholm University, SE-10691
              Stockholm, Sweden}

   \date{Received  / Accepted ?}

   \abstract{Solar-like differential rotation is characterized by a rapidly
     rotating equator and slower poles. 
     However, theoretical models and numerical simulations can also result
     in a slower equator and faster poles when the overall rotation is slow.
   }{%
     We study the critical rotational influence under which differential 
     rotation flips from solar-like (fast equator, slow poles) to an 
     anti-solar one (slow equator, fast poles).
     We also estimate the non-diffusive ($\Lambda$ effect) and diffusive 
     (turbulent viscosity) contributions to the Reynolds stress.
   }%
   {
     We present the results of three-dimensional numerical simulations of mildly turbulent
     convection in spherical wedge geometry. Here we apply a fully compressible 
     setup which would suffer from a prohibitive time step constraint if the 
     real solar luminosity was used.
     To avoid this problem while still representing the same rotational
     influence on the flow as in the Sun, we increase the luminosity by a
     factor of roughly $10^6$ and the rotation rate by a factor of $10^2$.
     We regulate the convective velocities by varying
     the amount of heat transported by thermal conduction, turbulent 
     diffusion, and resolved convection.
   }%
   {
     Increasing the efficiency of resolved convection leads to a
     reduction of the rotational influence on the flow and a sharp transition
     from solar-like to anti-solar differential rotation for Coriolis numbers
     around 1.3.
     We confirm the recent finding of a large-scale flow bistability:
     contrasted with running the models from an initial condition with
     unprescribed differential rotation, the initialization of the
     model with certain kind of rotation profile sustains the solution
     over a wider parameter range. 
     The anti-solar profiles are found to be more stable against
     perturbations in the level of convective turbulent velocity than
     the solar-type solutions.
   }%
   {
     Our results may have implications for real stars that start their lives
     as rapid rotators implying solar-like rotation in the early
     main-sequence evolution.  
     As they slow down, they might be able to retain solar-like
     rotation for lower Coriolis numbers, and thus longer in time, before
     switching to anti-solar rotation. This could partially explain the puzzling
     findings of anti-solar rotation profiles for models in the solar
     parameter regime.  }%


   \maketitle


\section{Introduction}

The solar surface differential rotation is characterized by a fast
equator and a monotonic decrease of angular velocity toward the poles.
The internal rotation of the Sun is such that only weak radial shear
is present in the bulk of the convection zone. Strong shear is present
only in the boundary layers at the bottom in the tachocline and in the
near-surface shear layer \cite[e.g.][and references
therein]{TCDMT03,MT09}. The differential rotation is explained by the
interaction of anisotropic turbulence and rotation, which is
represented by a non-diffusive contribution to the Reynolds stress
known as $\Lambda$ effect in mean-field hydrodynamics
\citep{KR74,R80,R89}. Furthermore, turbulent latitudinal heat fluxes, or
stably stratified layers below \citep{BMT11} or above \citep{WKMB13}
the convection zone are
needed to explain why the Taylor--Proudman balance is broken
in the Sun. Mean-field models with these ingredients or other
parameterizations have been successful in reproducing the solar
rotation profile \citep[e.g.][]{BMT92,KR95,Re05,HY11}, and recently also
the tachocline and the near-surface shear layer
\citep[e.g.][]{KO11}.

Mean-field models rarely produce anti-solar differential
rotation unless a strong meridional circulation is imposed
\citep{KR04}.
This suggestion is supported by simulations of \cite{DSB06}, which
displayed anti-solar differential rotation in their fully convective
models as a consequence of strong meridional circulation.
An arguably similar pathway was offered by \cite{AHW07}, who
proposed that strong mixing by turbulent convection would be
the primary agent for angular momentum equilibration and thus
anti-solar differential rotation.
This mixing must then be stronger in the meridional plane than
in the azimuthal direction to produce negative differential
rotation, similar to what is expected to occur in the near-surface
shear layer of the Sun \citep{KR05}.
Differential rotation and meridional circulation are intimately
coupled \citep{Kip63}, and one implies the other.
Anti-solar differential rotation implies a counter-clockwise circulation
in the northern hemisphere, corresponding to poleward motion at the surface.
In the context of the solar near-surface shear layer,
this mechanism is referred to as gyroscopic pumping \citep{MH11}.

Both mean-field models and simulations have limitations.
Furthermore, approximations like first-order smoothing are 
used to derive turbulent transport coefficients, such as turbulent 
viscosity and the so-called $\Lambda$ effect.
However, their range of validity in stellar convection 
zones is questionable and only limited comparisons between
hydrodynamic mean-field models and direct simulations have been 
performed \citep[see, however][]{RBMD94}.
Direct numerical simulations of convection, on the other hand,
lack small-scale structure and tend to be dominated by
structures that span the entire convection zone.

The rotational influence on the flow can be measured
by the local Coriolis number $\Co=2\Omega\tau$,
where $\Omega$ is the rotation rate and $\tau$ is the turnover time.
It is large in the bulk of the solar convection zone,
especially if $\tau$ is estimated from mixing length theory,
which predicts values of $\Co$ ranging from $10^{-3}$ near the surface
to more than $10$ in the deep layers \citep[e.g.][]{O03,BS05,K11}.
This is not captured by present simulations, which
might well be due to insufficient density stratification
in most numerical simulations so far.  Whether the deep layers of
stellar convection zones are really like this remains an open
question.  Observational evidence also points to anti-solar
differential rotation in some slowly rotating stars
\citep[e.g.][]{SKW03,WSW05} and solar-like rotation in rapidly
rotating dwarfs \citep[e.g.][]{CC02}.
However, for the star with the best observational evidence for
anti-solar differential rotation, namely the single K giant star
HD~31993 \citep{SKW03}, the Coriolis number is around unity and thus
close to the expected transition.

Recent numerical studies have explored the transition from anti-solar
to solar-like differential rotation
\citep{BP09,Chan10,KMGBC11,KMB11,GWA13,GYMRW14,GSKM13}. Here we
concentrate on the effect of superabiabaticity on the rotational
influence and resulting large-scale flows, and study the
bistability of the differential rotation first reported by
\cite{GYMRW14} using Boussinesq convection.
They found that, near the transition from anti-solar to solar-like
differential rotation, both kinds of solutions are possible -- depending
on the initial conditions. This is interesting for stellar
applications because most stars rotate rapidly when they are born and
slow down due to magnetic braking. Rapid rotation implies solar-like
differential rotation, which might then persist despite the angular
momentum loss due to stellar winds.

In this paper we set out to study the transition of solar-like
rotation profiles into the anti-solar regime, extending the work of
\citet{GYMRW14} into models of compressible convection. Using the
spherical wedge model employed in various previous papers, the
essential parts summarized in Section \ref{sect:model}, we investigate
the effect of changing the rotational influence by modifying the radiative
conductivity that has an effect on the convective velocities. We
perform two types of simulations, presented in
Section~\ref{sect:results}. Firstly, we run models from scratch,
i.e.\ without an initially prescribed rotation profile, and locate the
solar to
anti-solar transition in terms of Coriolis number. Secondly, we
investigate the dependence of the solutions on initial conditions by  
running models from solar and anti-solar states,
with otherwise identical parameters.

\section{The Model} \label{sect:model}

Our hydrodynamic model is essentially the same as the one used in
\cite{KMB11}. The same model has also been used to model
convection-driven dynamos \citep{KMB12,KMCWB13,CKMB14}. The
computational domain is a wedge in spherical polar coordinates, where
$(r,\theta,\phi)$ are radius, colatitude, and longitude. The
radial, latitudinal, and longitudinal extents of the wedge are $r_0
\leq r \leq r_1$, $\theta_0 \leq \theta \leq \pi-\theta_0$, and $0
\leq \phi \leq \phi_0$, respectively, where $r_0=0.72\,R_\odot$ and
$r_1=0.97\,R_\odot$ denote the positions of the bottom and top of the
computational domain, and $R_\odot=7\cdot10^8$m is the
radius of the Sun. Here we consider $\theta_0=\pi/12$ and
$\phi_0=\pi/2$, so we cover a quarter of the azimuthal extent between
$\pm75^\circ$ latitude.  
The dependence
on the latitudinal extent of the wedge was studied by \cite{KMGBC11},
who found that the results are robust as long as the
opening angle of the wedge is more than 90 degrees.
We solve the compressible hydrodynamic equations,
\begin{equation}
\frac{D \ln \rho}{Dt} = -\bm\nabla\cdot\bm{u},
\end{equation}
\begin{equation}
\frac{D\bm{u}}{Dt} = \bm{g} -2\bm\Omega_0\times\bm{u}+\frac{1}{\rho}
\left(\bm\nabla \cdot 2\nu\rho\bm{\mathsf{S}}-\bm\nabla p\right),
\end{equation}
\begin{equation}
T\frac{D s}{Dt} = -\frac{1}{\rho}\bm\nabla \cdot
\left({\bm F^{\rm rad}}+ {\bm F^{\rm SGS}}\right) +2\nu \bm{\mathsf{S}}^2,
\label{equ:ss}
\end{equation}
where $D/Dt = \pd/\pd t + \bm{u} \cdot \bm\nabla$ is the advective
time derivative, $\rho$ is the density, $\nu$ is the constant
kinematic viscosity,
\begin{equation}
{\bm F^{\rm rad}}=-K\ve{\nabla} T\quad\mbox{and}\quad
{\bm F^{\rm SGS}} =-\chiSGS \rho  T\ve{\nabla} s
\end{equation}
are radiative and subgrid-scale (hereafter SGS) heat fluxes, where $K$
is the radiative heat conductivity and $\chiSGS$ is the turbulent heat
conductivity, which represents the unresolved convective transport of
heat \citep{KMCWB13} and was referred to as $\chi_{\rm t}$ in
\cite{KMB11,KMB12}. Furthermore, $s$ is the specific entropy, $T$ is
the temperature, and $p$ is the pressure. The fluid obeys the ideal
gas law with $p=(\gamma-1)\rho e$, where $\gamma=c_{\rm P}/c_{\rm
  V}=5/3$ is the ratio of specific heats at constant pressure and
volume, respectively, and $e=c_{\rm V} T$ is the specific internal
energy. The rate of strain tensor $\bm{\mathsf{S}}$ is given by
\begin{equation}
\mathsf{S}_{ij}=\onehalf(u_{i;j}+u_{j;i})
-\onethird \delta_{ij}\bm\nabla\cdot\bm{u},
\end{equation}
where the semicolons denote covariant differentiation \citep{MTBM09}.

The gravitational acceleration is given by $\bm{g}=-GM_\odot\bm{r}/r^3$,
where $G=6.67\cdot10^{-11}$ m$^3$~kg$^{-1}$~s$^{-2}$ is the
gravitational constant, and $M_\odot=2.0\cdot10^{30}$ kg is the mass
of the Sun. We neglect self-gravity of the matter in the convection
zone. Furthermore, the rotation vector
$\bm\Omega_0$ is given by
$\bm\Omega_0=(\cos\theta,-\sin\theta,0)\Omega_0$.

\subsection{Initial and boundary conditions}
\label{sec:initcond}

Here we make an effort to connect the model more closely with the
parameters of the Sun.
Due to the fully compressible formulation of our model, we are faced
with a prohibitive time step limitation if we were to use the solar
luminosity.
As explained in detail in \cite{KMCWB13},
we circumvent this by using a roughly $10^6$ times higher
luminosity in the model in comparison to the Sun. As the convective
energy flux scales as $F_{\rm conv} \sim \rho u^3$, the convective
velocity $u$ is roughly 100 times greater in the simulations than in
the Sun.
To obtain the same rotational influence on the flow as in the Sun,
we must therefore increase $\Omega$ by the same factor. In general,
this can be written as
\begin{equation}
{\bm u}_{\rm sim}=L_{\rm ratio}^{1/3}{\bm u}_\odot \quad {\rm and}  \quad \Omega_{\rm sim}=L_{\rm ratio}^{1/3} \Omega_\odot,
\end{equation}
where $L_{\rm ratio}=L_0/L_\odot$, with $L_0$ and
$L_\odot\approx3.84\cdot10^{26}$~W being the luminosities of the
model and the Sun, respectively, and
$\Omega_\odot\approx2.7\cdot10^{-6}$s$^{-1}$ is the mean solar
rotation rate, corresponding to $430$~nHz. In what follows we scale
our results back to solar
units so that, say for the velocity, we quote ${\bm u}_{\rm
  sim}/L_{\rm ratio}^{1/3}$. 
The scaling used here is based on dimensional arguments.
It is supported by mixing length theory \citep{Vit53} and simulations
\citep{BCNS05,MFRT12}, and should be applicable as long as the energy transport
is not yet affected by rotation \citep[see e.g.][]{YGCD13}.
Furthermore, we assume that the density
and the temperature at the base of the convection zone at
$r=0.72R_\odot$ have the solar values $\rho_0=200$ kg~m$^{-3}$ and
$T_0=2.23\cdot10^6$K.

\begin{figure*}[t]
\centering
\includegraphics[width=\textwidth]{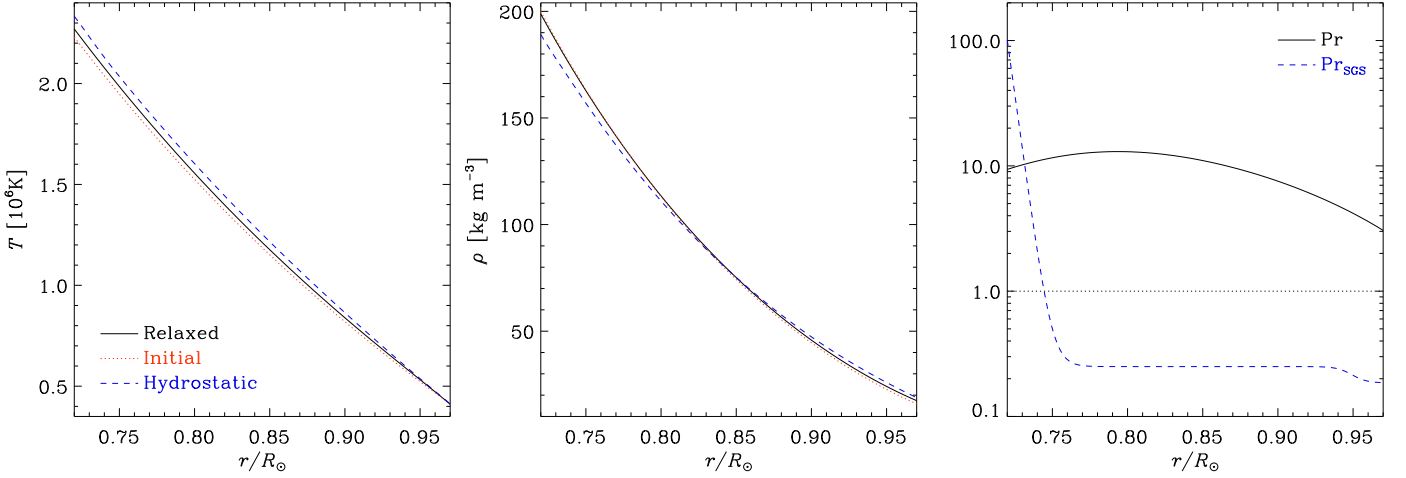}
\caption{Left and middle panels: temperature $T$ in units of $10^6\K$
  and density $\rho$ in units of $\kg\m^{-3}$, respectively, for the
  thermally relaxed state (black solid line), initial condition (red
  dotted), and hydrostatic solutions (blue dashed). Rightmost panel:
  Prandtl numbers related to radiative diffusion $\Pra=\nu/\chi$
  where $\chi=K/\rho c_{\rm P}$ (black solid line), and the turbulent
  heat conductivity $\PraSGS=\nu/\chi_{\rm SGS}$ (blue dashed) as
  functions of radius. Data is taken from Run~D.}
\label{fig:pstrat}
\end{figure*}

The initial state is isentropic and the hydrostatic temperature
gradient is given by
\begin{equation}
\frac{\pd T}{\pd r}=-\frac{GM_\odot/r^2}{c_{\rm V}(\gamma-1)(n_{\rm ad}+1)},
\end{equation}
where $n_{\rm ad}=1.5$ is the polytropic index for an adiabatic stratification.
We fix the value of $\pd T/\pd r$ on the lower boundary.
The density profile follows from hydrostatic equilibrium.
To speed up the thermal relaxation, the initial condition is chosen
not to be in thermodynamic equilibrium, but closer to the final
convecting state.
We choose the heat conduction profile such that
radiative diffusion is responsible for supplying the energy flux in
the system and progressively less so further out by choosing
a radiative conductivity, $K(r)=K_0[n(r)+1]$, with
$n(r)=\delta n (r/r_0)^{-15}+n_{\rm ad}- \delta n$ replacing the 
polytropic index,
\begin{equation}
K_0=(\mathcal{L}/4\pi)c_{\rm V}(\gamma-1)(n_{\rm ad}+1)\rho_0\sqrt{GMR},
\end{equation}
being a reference conductivity, and $\mathcal{L}$ being the non-dimensional
luminosity, given below. Now $n=n_{\rm ad}$ at the bottom of the
convection zone and approaches $n_{\rm ad} -\delta n$ at the surface.
This means that $K=(n+1)K_0$ decreases toward the surface like
$r^{-15}$ such that the value of $\delta n$ regulates the flux that is
carried by convection \citep{BCNS05}.
Initial, final, and hydrostatic profiles of the temperature and
density as well as the profiles of $\PraSGS=\nu/\chiSGS$ and
$\Pra=\nu/\chi$, where $\chi=K/\rho c_{\rm P}$, are shown in
Fig.~\ref{fig:pstrat}.
We introduce weak small-scale Gaussian noise velocity perturbations in
the initial state.

Our simulations are defined by the energy flux imposed at the bottom
boundary, $F_{\rm b}=-(K \pd T/\pd r)|_{r=r_0}$
as well as the values of $\Omega_0$, $\nu$, and
$\chiSGSm=\chiSGS(r_{\rm m}=0.845\, R_\odot)$.
Furthermore, the radial profile of $\chiSGS$ is piecewise constant above
$r>0.75R_\odot$ with $\chiSGS=\chiSGSm$ at $0.75R_\odot < r <0.95R_\odot$, and
$\chiSGS=1.35\chiSGSm$ above $r=0.95R_\odot$. Below $r=0.75R_\odot$,
$\chiSGS$ tends smoothly to zero; see Fig.~1 of \cite{KMB11}.
We fix the value of $\chiSGS$ such that it corresponds to
$5\cdot10^8\m^2\s^{-1}$ in physical units at $r=r_1$.

The radial and latitudinal boundaries are assumed to be impenetrable and
stress free, i.e.,
\begin{eqnarray}
&&\!\!\!
u_r=0,\quad \frac{\pd u_\theta}{\pd r}=\frac{u_\theta}{r},\quad \frac{\pd
u_\phi}{\pd r}=\frac{u_\phi}{r} \quad (r=r_0, r_1),\\
&&\!\!\!
\frac{\pd u_r}{\pd \theta}=u_\theta=0,\quad \frac{\pd u_\phi}{\pd
\theta}=u_\phi \cot \theta \quad (\theta=\theta_0,\pi-\theta_0).
\quad
\end{eqnarray}
Density and specific entropy have vanishing first derivatives on the
latitudinal boundaries, thus suppressing heat fluxes through
them.

On the outer radial boundary we apply a black body condition
\begin{equation}
\sigma T^4  = -K\nabla_r T - \chiSGS \rho T \nabla_r s,
\label{eq:bbb}
\end{equation}
where $\sigma$ is the Stefan--Boltzmann constant. We use a modified
value for $\sigma$ that takes into account that both surface
temperature and energy flux through the domain are larger than in the
Sun. We choose $\sigma$ such that the flux at the surface, $\sigma T^4$, 
carries the total luminosity through the boundary in the initial
non-convecting state.

\begin{table*}[t!]
\centering
\caption[]{Summary of the runs.}
      \label{tab:runs}
      \vspace{-0.5cm}
     $$
         \begin{array}{p{0.05\linewidth}ccccccrrcccccccc}
           \hline
           \noalign{\smallskip}
Run & \Ra  & \Pra & \delta n  & \Roc & \Rey & \Co & \Delta_\Omega^{(\theta)} & \Delta_\Omega^{(r)} & \tilde{E}_{\rm kin}[10^{-7}] &
E_{\rm mer}/E_{\rm kin} & E_{\rm rot}/E_{\rm kin} & \tilde{L}_{\rm rad} & \tilde{L}_{\rm conv} & {\rm remarks} \\ \hline 
A  & 3.93\cdot10^5 & 40.4 & 2.50 & 0.73 & 35 & 1.24 & -2.47 & -0.51 & 4.20 & 0.003 & 0.954 & 0.09 & 1.01 & {\rm AS} \\ 
B  & 3.54\cdot10^5 & 20.3 & 2.25 & 0.69 & 34 & 1.27 & -2.36 & -0.50 & 3.87 & 0.003 & 0.954 & 0.18 & 0.84 & {\rm AS} \\ 
C  & 3.16\cdot10^5 & 13.6 & 2.00 & 0.65 & 33 & 1.33 & -2.28 & -0.49 & 3.60 & 0.003 & 0.955 & 0.30 & 0.66 & {\rm AS} \\ 
D  & 2.92\cdot10^5 & 11.3 & 1.85 & 0.63 & 29 & 1.50 &  0.03 &  0.04 & 0.33 & 0.001 & 0.466 & 0.34 & 0.62 & {\rm SL} \\ 
E  & 2.77\cdot10^5 & 10.1 & 1.75 & 0.61 & 27 & 1.60 & -0.01 &  0.09 & 0.38 & 0.001 & 0.593 & 0.37 & 0.53 & \mbox{SL, polar jet} \\ 
           \hline
D0 & 2.92\cdot10^5 & 11.3 & 1.85 & 0.63 & 29 & 1.50 &  0.03 &  0.04 & 0.33 & 0.001 & 0.466 & 0.34 & 0.62 & {\rm SL} \\ 
D1 & 3.16\cdot10^5 & 13.5 & 2.00 & 0.65 & 29 & 1.50 &  0.16 &  0.07 & 0.37 & 0.001 & 0.517 & 0.28 & 0.65 & {\rm SL} \\ 
D2 & 3.31\cdot10^5 & 15.6 & 2.10 & 0.67 & 30 & 1.44 &  0.13 &  0.05 & 0.37 & 0.001 & 0.465 & 0.24 & 0.72 & {\rm SL} \\ 
D3 & 3.47\cdot10^5 & 18.4 & 2.20 & 0.68 & 32 & 1.37 &  0.03 &  0.01 & 0.28 & 0.002 & 0.243 & 0.21 & 0.81 & {\rm SL} \\ 
D4 & 3.62\cdot10^5 & 22.7 & 2.30 & 0.70 & 35 & 1.27 & -2.33 & -0.50 & 3.93 & 0.003 & 0.955 & 0.19 & 0.87 & {\rm AS} \\ 
           \hline
B0 & 3.54\cdot10^5 & 20.3 & 2.25 & 0.69 & 34 & 1.27 & -2.36 & -0.50 & 3.87 & 0.003 & 0.954 & 0.18 & 0.84 & {\rm AS} \\ 
B1 & 3.47\cdot10^5 & 18.5 & 2.20 & 0.68 & 34 & 1.27 & -2.26 & -0.50 & 3.92 & 0.002 & 0.956 & 0.24 & 0.85 & {\rm AS} \\ 
B2 & 3.31\cdot10^5 & 15.7 & 2.10 & 0.67 & 33 & 1.31 & -2.27 & -0.50 & 3.87 & 0.003 & 0.958 & 0.27 & 0.77 & {\rm AS} \\ 
B3 & 3.16\cdot10^5 & 13.6 & 2.00 & 0.65 & 33 & 1.34 & -2.18 & -0.49 & 3.67 & 0.003 & 0.958 & 0.31 & 0.71 & {\rm AS} \\ 
B4 & 3.00\cdot10^5 & 12.0 & 1.90 & 0.64 & 32 & 1.37 & -2.15 & -0.49 & 3.54 & 0.003 & 0.958 & 0.34 & 0.64 & {\rm AS} \\ 
B5 & 2.85\cdot10^5 & 10.8 & 1.80 & 0.62 & 31 & 1.41 & -2.06 & -0.49 & 3.38 & 0.003 & 0.958 & 0.38 & 0.59 & {\rm AS} \\ 
B6 & 2.69\cdot10^5 &  9.7 & 1.70 & 0.60 & 30 & 1.45 & -1.98 & -0.48 & 3.16 & 0.003 & 0.957 & 0.41 & 0.54 & {\rm AS} \\ 
B7 & 2.38\cdot10^5 &  8.2 & 1.50 & 0.57 & 28 & 1.54 & -1.86 & -0.47 & 2.83 & 0.003 & 0.958 & 0.48 & 0.44 & {\rm AS} \\ 
B8 & 2.22\cdot10^5 &  7.6 & 1.40 & 0.55 & 27 & 1.60 & -1.76 & -0.45 & 2.63 & 0.003 & 0.958 & 0.51 & 0.39 & {\rm AS} \\ 
B9 & 2.07\cdot10^5 &  7.0 & 1.30 & 0.53 & 22 & 2.01 &  0.04 &  0.12 & 0.25 & 0.001 & 0.632 & 0.54 & 0.28 & \mbox{SL, polar jets} \\ 
B10 & 1.59\cdot10^5 & 5.8 & 1.00 & 0.46 & 18 & 2.43 &  0.06 &  0.12 & 0.19 & 0.001 & 0.668 & 0.65 & 0.17 & \mbox{SL, polar jets} \\ 
B10b & 1.59\cdot10^5 & 5.8 & 1.00 & 0.46 & 17 & 2.61 &  0.33 &  0.11 & 0.34 & 0.000 & 0.831 & 0.65 & 0.17 & {\rm SL} \\ 
           \hline
         \end{array}
     $$
\tablefoot{
All models have $\Pra_{\rm SGS}=0.25$, $\mathcal{L}=3.85\cdot10^{-5}$, 
$\Ta=2.98\cdot 10^{6}$, $\xi=0.0325$ corresponding to $\Gamma\approx 12$,
$\Omega_0/\Omega_\odot=1$, 
and use a grid resolution $128\times256\times128$.
With $\chiSGS(r_1)=5\cdot10^8\m^2\s^{-1}$, we have
$\chiSGSm=3.7\cdot10^8\m^2\s^{-1}$, and
$\nu=9.3\cdot10^7\m^2\s^{-1}$.
$\tilde{E}_{\rm kin}=\brac{\onehalf \rho \bm{u}^2}$ is the volume averaged
total kinetic energy, in units of $GM_\odot\rho_0/R_\odot$.
$E_{\rm mer}=\onehalf\brac{\rho(\mean{u}_r^2+\mean{u}_\theta^2)}$ 
and $E_{\rm rot}=\onehalf\brac{\rho\mean{u}_\phi^2}$ are the kinetic
energies of the meridional circulation and differential rotation.
$\tilde{L}_{\rm rad}$ and $\tilde{L}_{\rm conv}$ are the fractions of total 
flux transported by radiative conduction and resolved convection at 
$r=r_{\rm m}$.
Runs~D and D0, and Runs~B and B0 are the same. Run~D1 was continued from
a snapshot of Run~D, whereas the other models in Set~D 
were continued from a relaxed state of D1. In Set~B, a snapshot
of Run~B0 was used as an initial condition.
In the last column, AS and SL stand for anti-solar and solar-like 
differential rotation, respectively.}
\end{table*}

\subsection{Dimensionless parameters}
\label{sec:dimless}

The non-dimensional input parameters of our models are the luminosity
parameter
\begin{equation}
\mathcal{L} = \frac{L_0}{\rho_0 (GM_\odot)^{3/2} R_\odot^{1/2}},
\end{equation}
the normalized pressure scale height at the surface,
\begin{equation}
\xi = \frac{(\gamma-1) c_{\rm V}T_1}{GM_\odot/R_\odot},
\end{equation}
with $T_1$ being the temperature at the surface,
the Taylor number
\begin{equation}
\Ta=(2\Omega_0 \Delta r^2/\nu)^2,
\end{equation}
where $\Delta r=r_1-r_0=0.25\,R_\odot$, as well as the
fluid and SGS Prandtl numbers
\begin{equation}
\Pra=\frac{\nu}{\chi_{\rm m}},\quad \Pra_{\rm SGS}=\frac{\nu}{\chiSGSm},
\end{equation}
where $\chi_{\rm m}=K(r_{\rm m})/c_{\rm P} \rho_{\rm m}$
is the thermal diffusivity and $\rho_{\rm m}$ is the density,
both evaluated at $r=r_{\rm m}=0.845R_\odot$.
We vary $\Pra$ and keep $\Pra_{\rm SGS}=0.25$ fixed.
Finally, the non-dimensional viscosity is
\begin{equation}
\tilde{\nu}=\frac{\nu}{\sqrt{GM_\odot R_\odot}}.
\end{equation}
In addition to $\xi$, we quote the initial density contrast,
$\Gamma_\rho^{(0)}\equiv\rho(r_0)/\rho(r_1)$.
In the current moderately stratified simulations the density contrast
changes by less than 10 per cent during the run, see the middle panel of
Fig.~\ref{fig:pstrat}.

All other parameters are used as diagnostics and are not input
parameters. These include the fluid Reynolds number
\begin{equation}
\Rey=\frac{\urms}{\nu \kef},
\end{equation}
where $\kef=2\pi/\Delta r\approx25 R_\odot^{-1}$ is an estimate of the
wavenumber of the largest eddies. The Coriolis number is defined as
\begin{equation}
\Co=\frac{2\Omega_0}{\urms \kef},
\label{eq:Coriolis}
\end{equation}
where $\urms=\sqrt{(3/2)\brac{u_r^2+u_\theta^2}_{r\theta\phi t}}$ is
the rms velocity and the subscripts indicate averaging over $r$,
$\theta$, $\phi$, and a time interval during which the run is
thermally relaxed.
For $\urms$ we omit the
contribution from the azimuthal velocity, because it is
dominated by the differential rotation \citep{KMGBC11}.
To have a reasonable estimate of the rms velocity similar to that under
isotropic conditions, we compensate for the omission of $u_\phi^2$
by the $3/2$ factor. The Taylor number can
also be written as $\Ta=\Co^2\Rey^2(\kef R_\odot)^4$.
Furthermore, we define the Rayleigh number as
\begin{eqnarray}
\Ra\!=\!\frac{GM_\odot(\Delta r)^4}{\nu \chiSGSm R_\odot^2} \bigg(-\frac{1}{c_{\rm P}}\frac{{\rm d} s_{\rm hs}}{{\rm d}r} \bigg)_{r_{\rm m}},
\label{equ:Co}
\end{eqnarray}
where $s_{\rm hs}$ is the entropy in the hydrostatic, non-convecting
state. We compute the hydrostatic stratification by evolving a
one-dimensional model (no convection) with the values and profiles of $K$ and $\chiSGS$
given above. We also quote the convective Rossby number
  \citep{Gi77}
\begin{equation}
\Roc = \left( \frac{\Ra}{\Pra_{\rm SGS} \Ta}  \right)^{1/2}.
\end{equation}
We define mean quantities as averages over the
$\phi$-coordinate and denote them by overbars.
We also often average the data in time over the period of the
  simulations where thermal energy and differential rotation have
  reached statistically saturated states.

The simulations are performed with the {\sc Pencil
  Code}\footnote{http://pencil-code.google.com/}, which uses a
high-order finite difference method for solving the compressible
equations of magnetohydrodynamics.

\begin{figure}[t]
\centering
\includegraphics[width=\columnwidth]{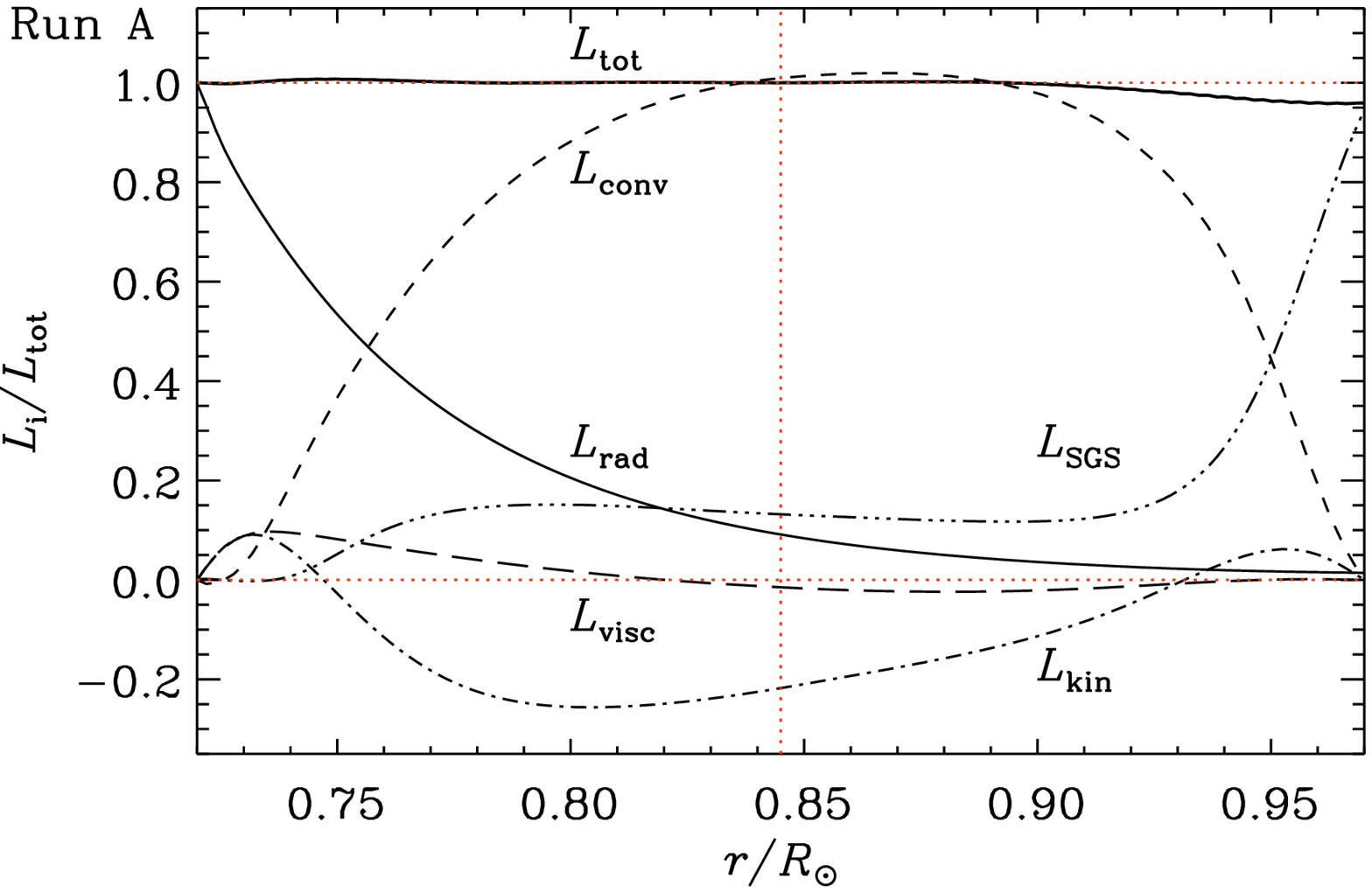}
\includegraphics[width=\columnwidth]{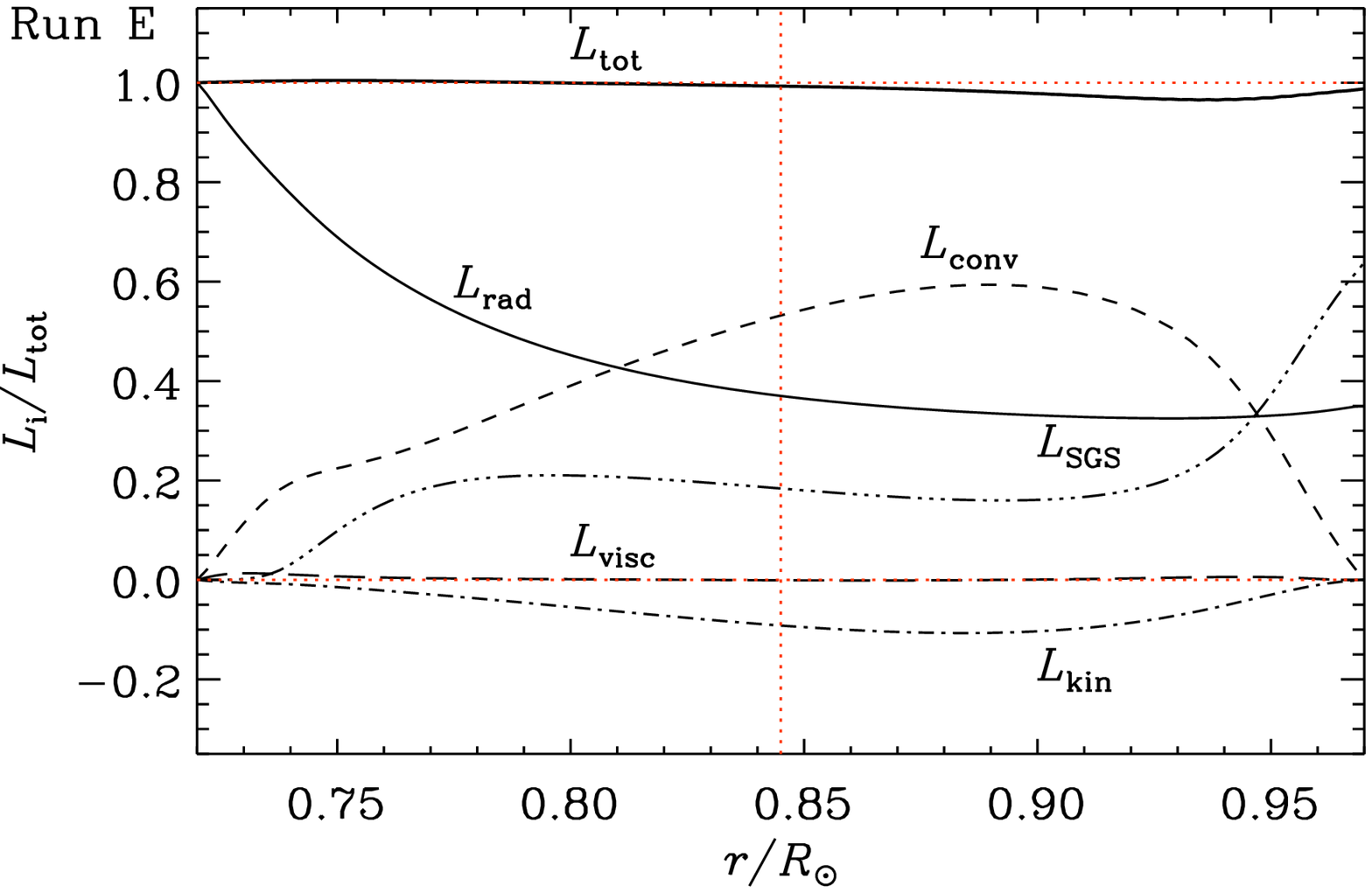}
\caption{Time-averaged energy fluxes from Runs~A (top) and E 
(bottom): radiative 
  (thin solid line),
  convective (dashed), kinetic energy (dash-dotted), SGS
  (triple-dash-dotted), and viscous (long-dashed) flux. The thick
  solid line denotes the total flux, whereas the red horizontal dotted
  lines show the zero and unity line. The red vertical dotted line at
  $r=r_{\rm m}$ shows the midpoint of the convection zone.}
\label{fig:pflu_A}
\end{figure}

\begin{figure}[t]
\centering
\includegraphics[width=\columnwidth]{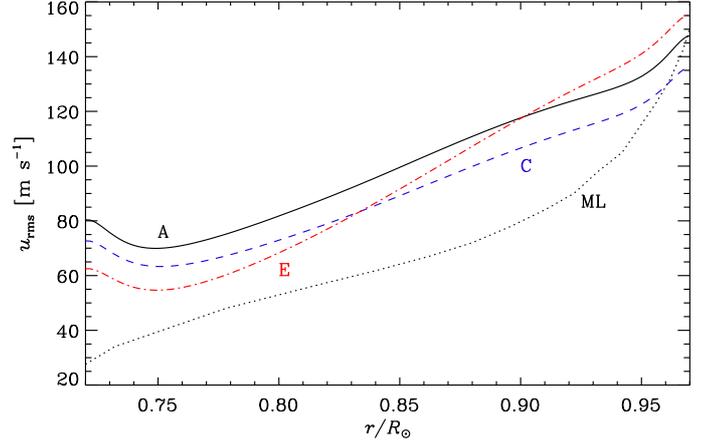}
\caption{Radial dependence of the time averaged fluctuating 
  rms-velocity where the
  contributions from the mean flows are omitted for Runs~A (solid
  black line), C (blue dashed), and E (red dot-dashed). The black
  dotted line shows the convective velocity from the mixing length (ML)
  model of \cite{Stix02}.}
\label{fig:purmsr}
\end{figure}

\section{Results} \label{sect:results}

Our simulations are summarized in Table~\ref{tab:runs}. We perform three
sets of runs. In the first set, we run the model from the initial
conditions described in Sect.~\ref{sec:initcond}
(Runs~A--E).
Here, Runs~A--C turn out to have anti-solar differential rotation,
while Runs~D and E have solar-like differential rotation.
Secondly, we study the bistability of the rotation
profile by taking either a solar-like (Runs~D0--D4) or an anti-solar
(Runs~B0--B10b) solution as initial conditions. Apart from $\delta
n$, we keep all other parameters fixed. We also estimate
$\Lambda$ effect coefficients from the simulation results.

\subsection{Effect of varying radiative flux}

We change the radiative conductivity by varying the
parameter $\delta n$, which regulates the amount of flux that
convection has to transport, thus influencing the convective
velocities and the Coriolis number. The different contributions to the
total energy flux from Runs~A and E are shown in
Fig.~\ref{fig:pflu_A}. The definitions of the fluxes can be found in
\cite{KMCWB13}. We give the fractions of convective and radiative
contributions to the total energy flux at the middle of the convection
zone in
Table~\ref{tab:runs}. For $\delta n=2.5$ (Run~A) the convective flux
can exceed the total flux, so the radiative flux transports less than
10 per cent of the luminosity in the upper part of the convection
zone. In Run~E with $\delta n = 1.75$ the fractions of radiative
diffusion and convection are 37 and 53 per cent, respectively. In the
extreme case of $\delta n=1$ (Runs~B10 and B10b),
convection transports only about 20 per cent of the flux. These 
cases are comparable to the setups used in earlier works
\citep{KKBMT10,KMGBC11}.

\begin{figure*}[t]
\centering
\includegraphics[width=0.2\textwidth]{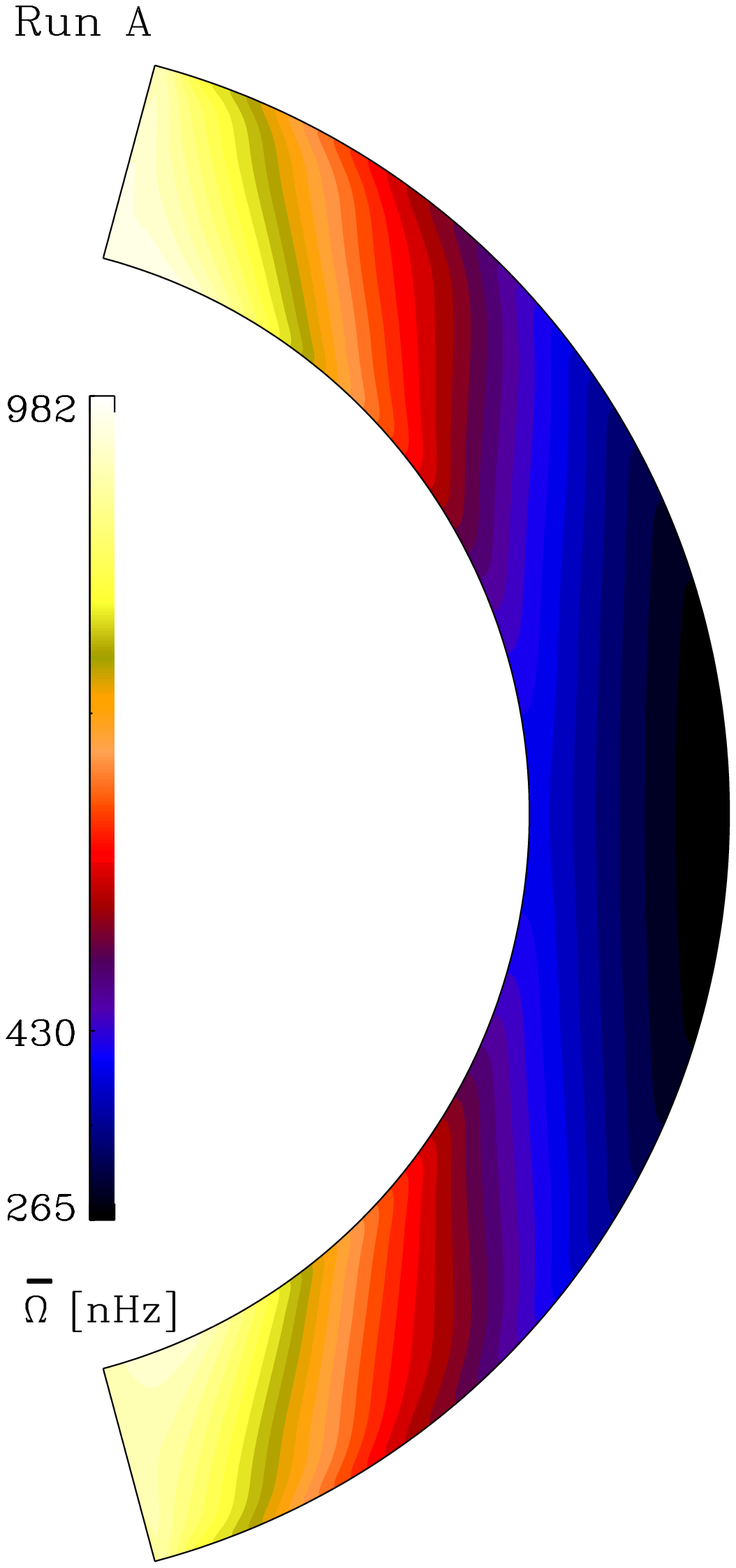}\includegraphics[width=0.2\textwidth]{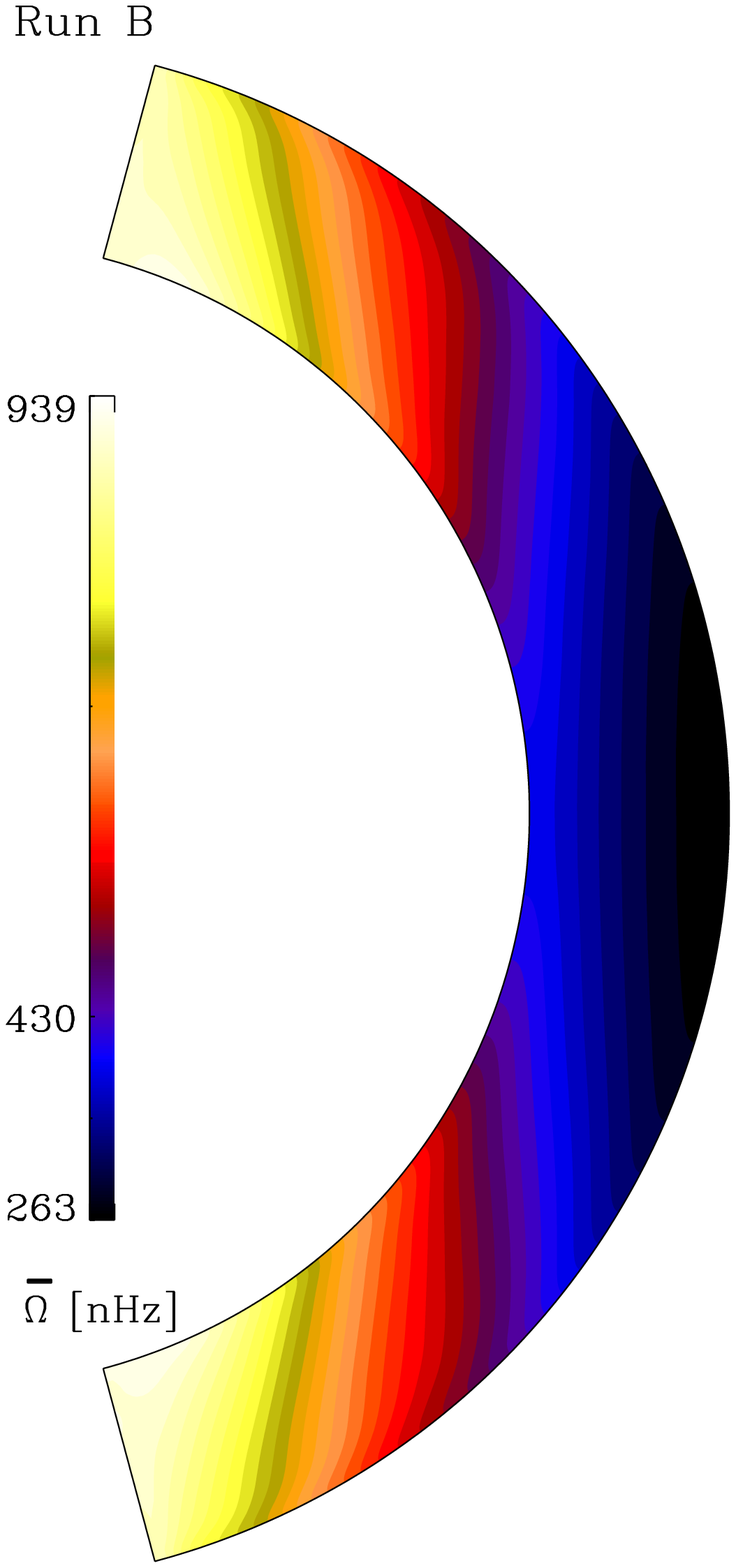}\includegraphics[width=0.2\textwidth]{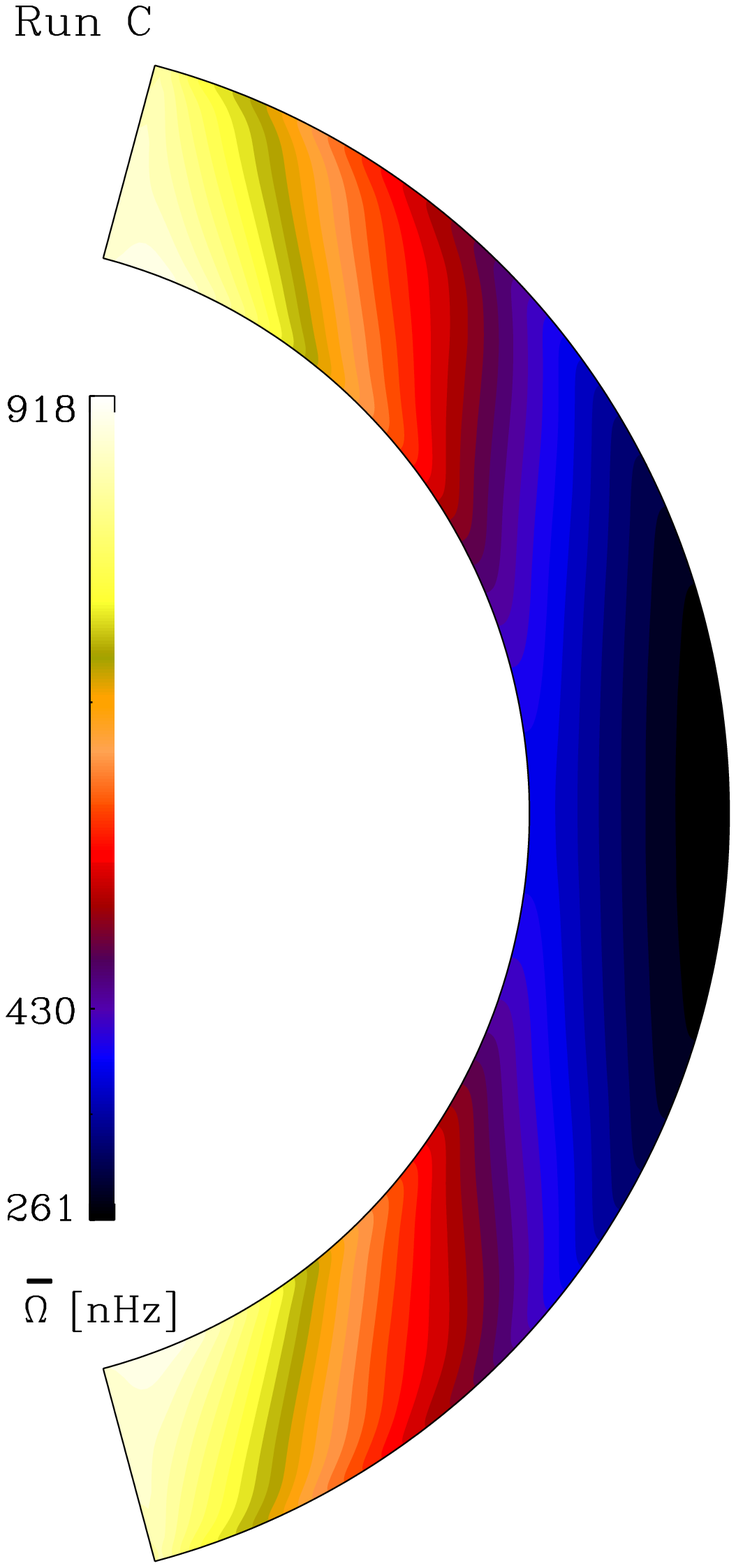}\includegraphics[width=0.2\textwidth]{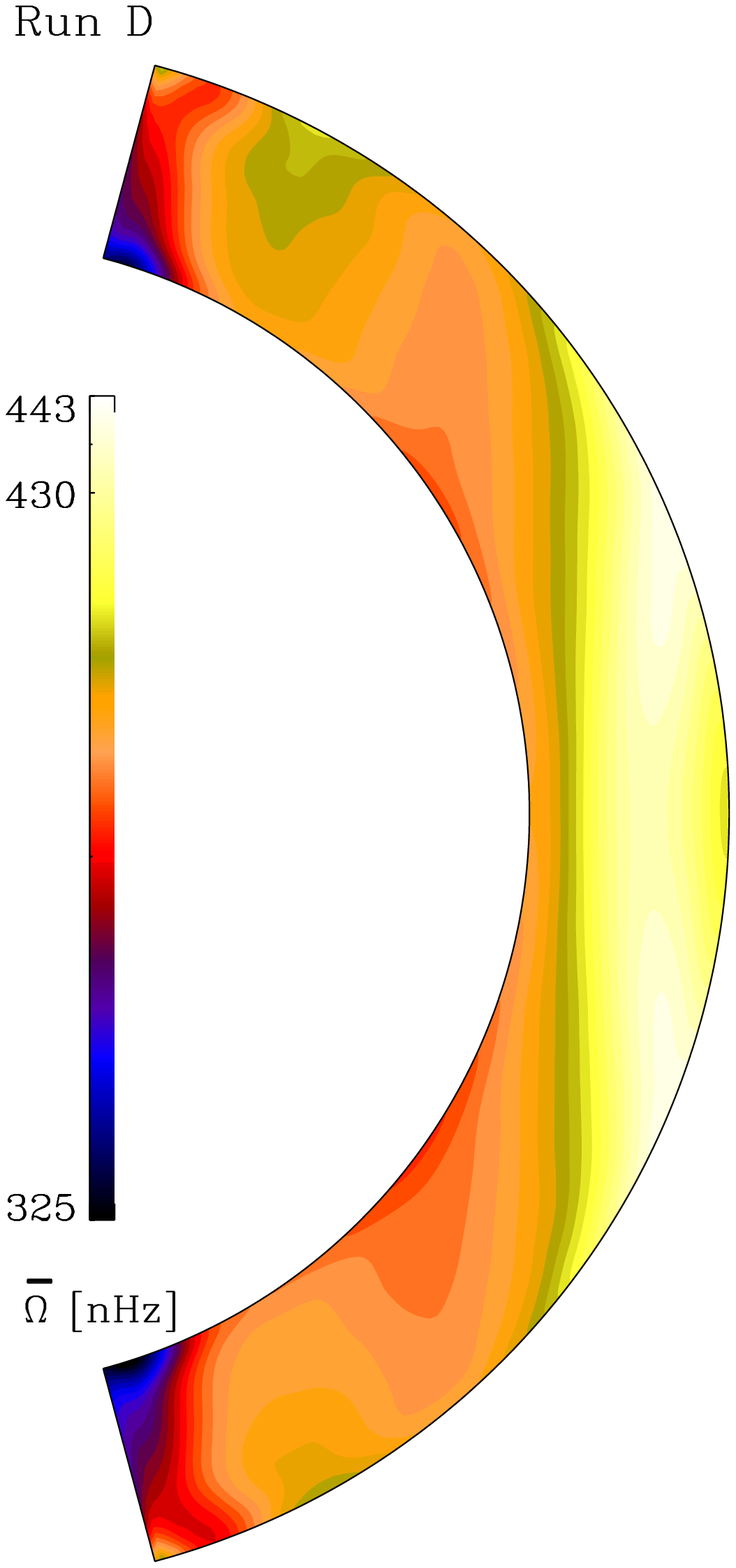}\includegraphics[width=0.2\textwidth]{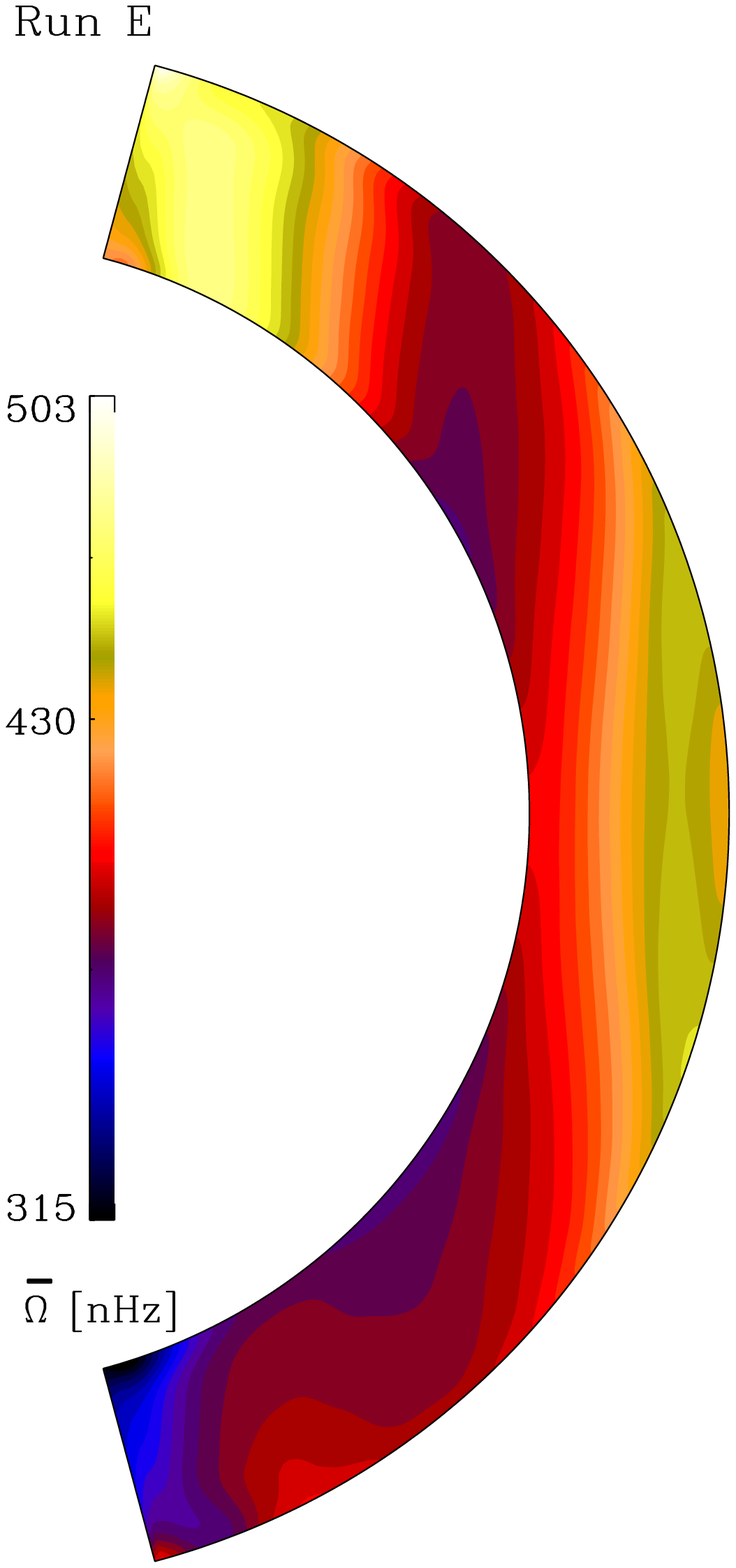}
\caption{Time-averaged rotation profiles from Runs~A--E showing $\mean\Omega$ in nHz.}
\label{fig:pOm}
\end{figure*}

The rms-velocities based on the
fluctuating velocity for Runs~A, C, and E are shown in
Fig.~\ref{fig:purmsr}. The changes between Runs~A and C are rather
subtle, the main effect being the decrease in the overall magnitude
of $\urms$. The main difference between Runs~C and E is the
increased contrast of the values near the boundaries. We also
find that all runs 
show higher velocities than the mixing length (ML) model of
  \cite{Stix02} below $r=0.95R_\odot$.
  The increase in $\urms$ near the lower boundary is due to our 
  impenetrable boundary condition. 
  Near the upper boundary our values of $\urms$ are close to
  those obtained from ML. A likely explanation is the weaker
  density stratification used in our simulations ($\Gamma\approx 12$)
  as opposed to what is realized in the ML model ($\Gamma\approx 60$), 
  leading to higher values near the surface in the latter.
Computing an average velocity using the data in
  Fig.~\ref{fig:purmsr} and using that to compute a Coriolis number
  according to Eq.~(\ref{eq:Coriolis}), we find $\Co\approx2.13$ for
  the ML model. This is clearly above the values for our Runs~A--E
  owing to the lower velocities.
However, we note that recent high-resolution simulations of 
  non-rotating
  solar convection suggest significantly higher velocities than
  anticipated from ML models \citep{HRY14} so the actual Coriolis
  number of the Sun might also be much smaller than our estimate.

\begin{figure*}[t]
\centering
\includegraphics[width=0.4\textwidth]{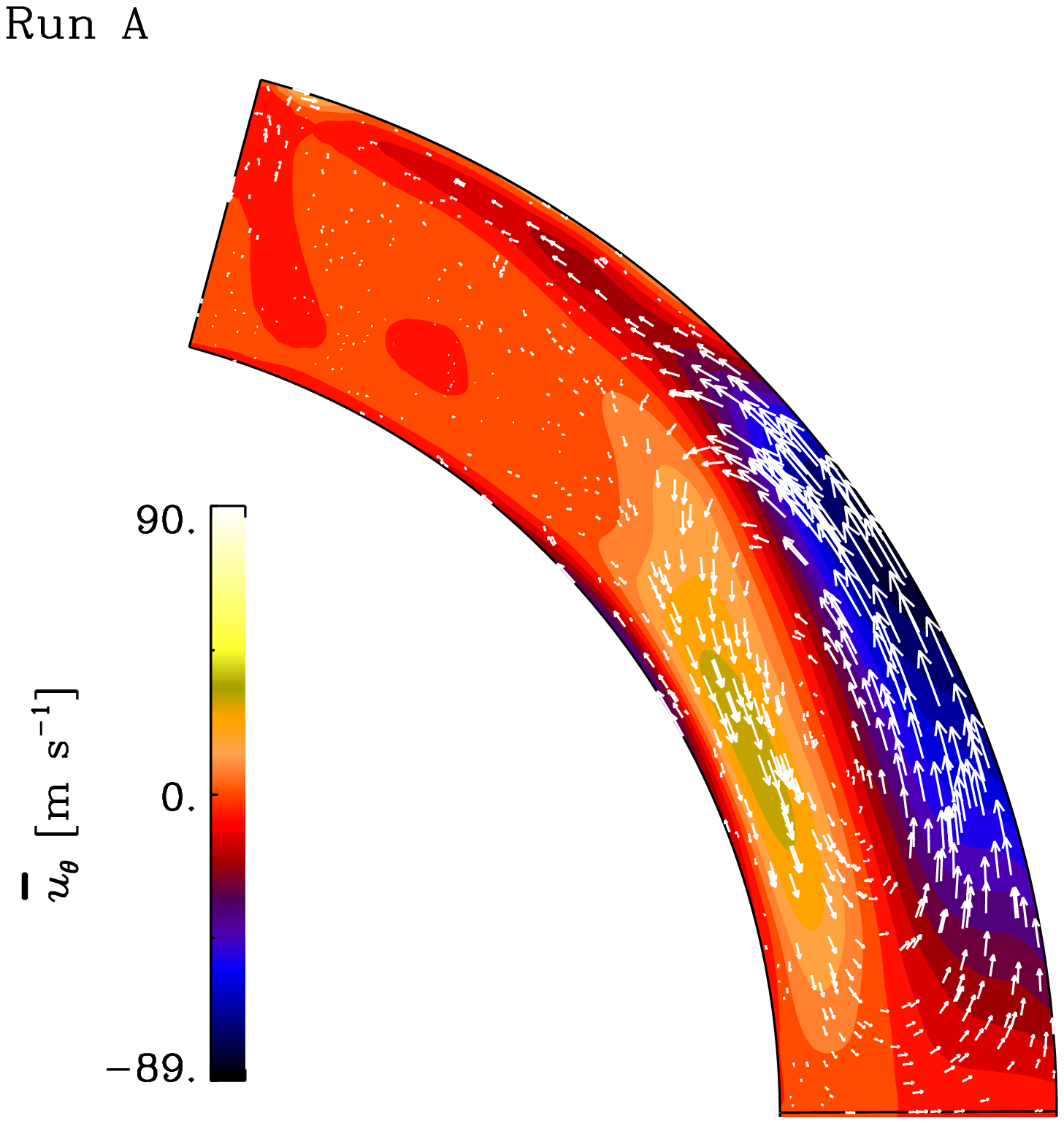}\includegraphics[width=0.4\textwidth]{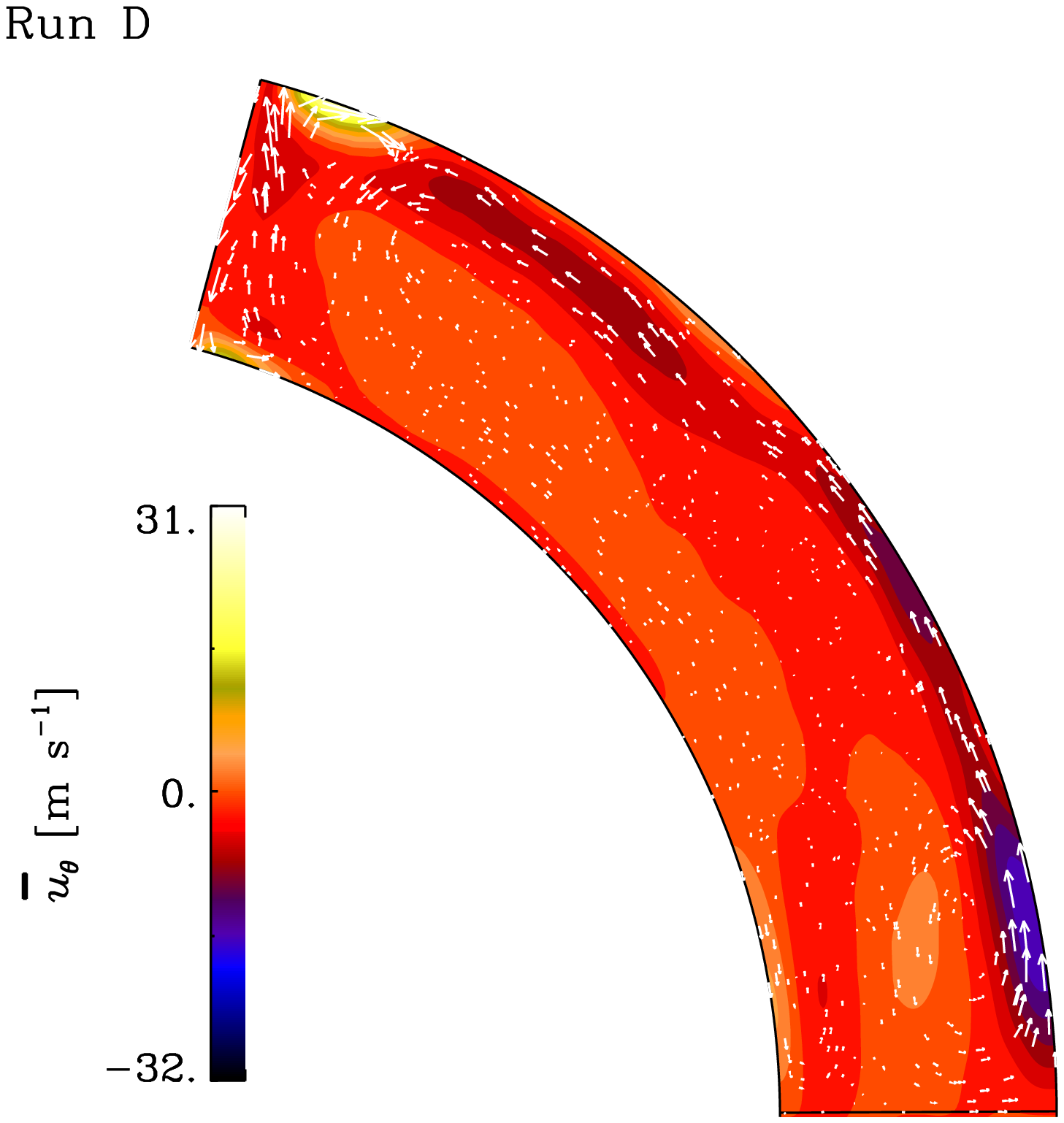}
\caption{Time-averaged meridional circulation from Runs~A (left) and D
  (right). The arrows show the flow $\mean{\bm u}_{\rm
    m}=(\mean{u}_r,\mean{u}_\theta)$, whereas the colour contours show
  $\mean{u}_\theta$.}
\label{fig:pMC}
\end{figure*}

\begin{figure*}[t]
\centering
\includegraphics[width=0.5\textwidth]{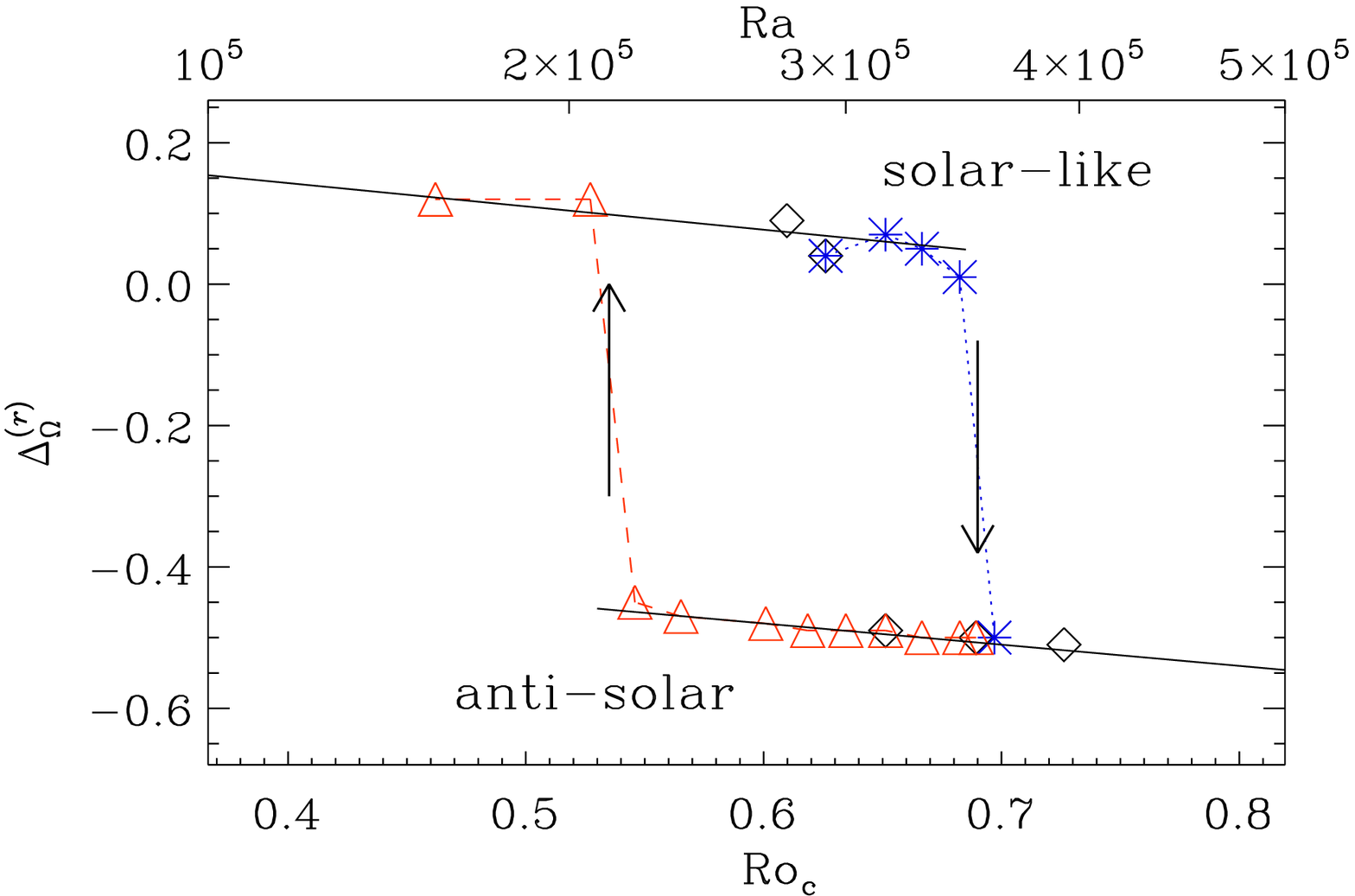}\includegraphics[width=0.5\textwidth]{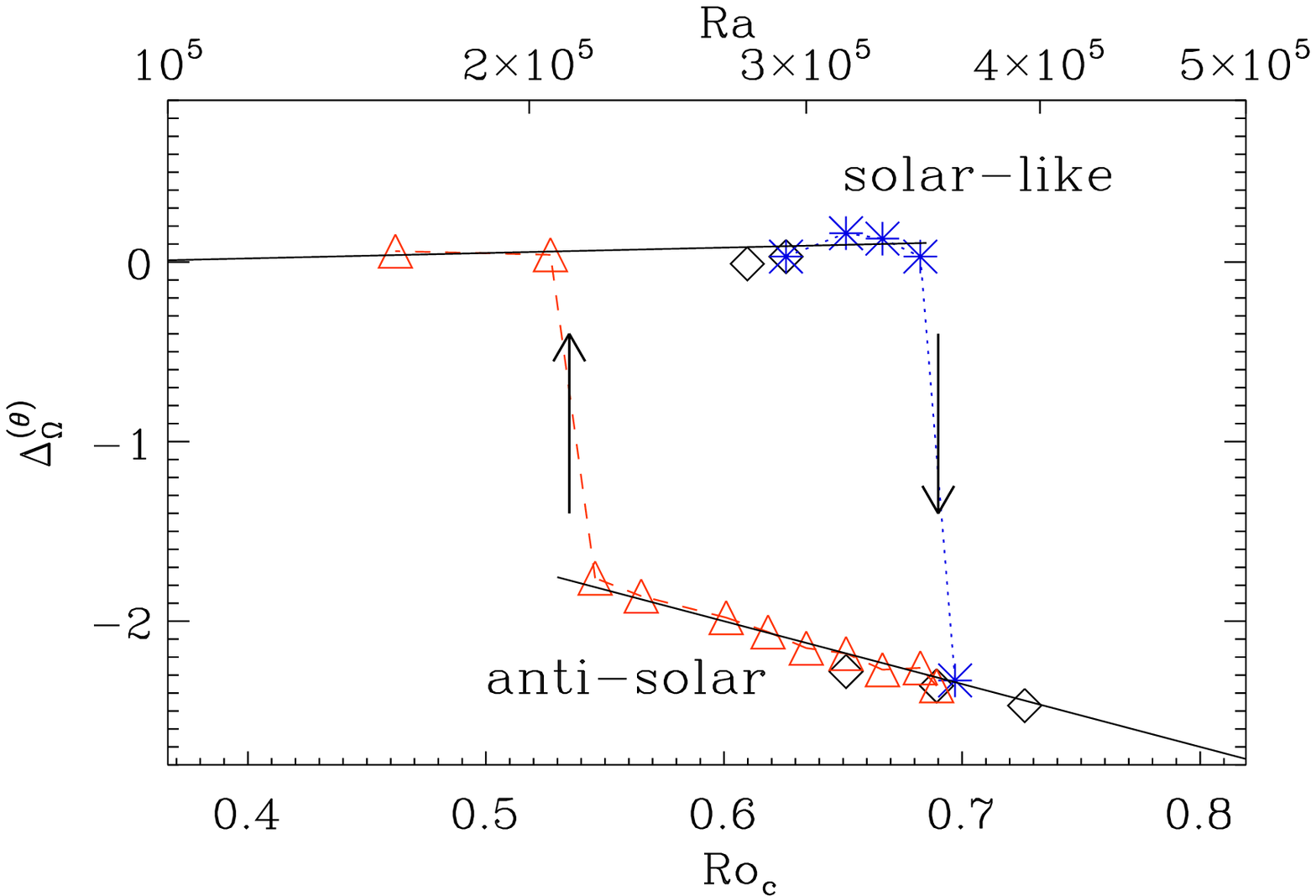}
\caption{Radial (left panel) and latitudinal (right panel)
  differential rotation, defined by Eqs.~(\ref{equ:pDRt}),
  from Runs~A--E (diamonds), and Set~D (blue dotted line with asterisks)
  and B (red dashed line with triangles).}
\label{fig:pDR}
\end{figure*}

\subsection{Differential rotation}

The decrease in the rms-velocity, since $\delta n$ is lowered, is also reflected
in the Coriolis number, which more than doubles between the extreme
cases A and B10. The increasing rotational effect on the flow affects
the rotation profile realized in the runs.
The profiles of $\mean\Omega=\mean{u}_\phi/r \sin \theta + \Omega_0$
are shown in Fig.~\ref{fig:pOm} for Runs~A--E,
which were run from the initial conditions stated in
Sect.~\ref{sec:initcond} with $\delta n$ varying systematically.
We characterize the relative radial and latitudinal differential rotation
by the quantities \citep{KMCWB13}
\begin{eqnarray}
\Delta_\Omega^{(r)}=\frac{\Omega_{\rm eq}-\Omega_{\rm bot}}{\Omega_{\rm eq}},
\quad
\Delta_\Omega^{(\theta)}=\frac{\Omega_{\rm eq}-\Omega_{\rm pole}}{\Omega_{\rm eq}},\label{equ:pDRt}
\end{eqnarray}
where $\Omega_{\rm eq}=\mean\Omega(r_1,\pi/2)$ and $\Omega_{\rm
  bot}=\mean\Omega(r_0,\pi/2)$ are the equatorial rotation rates at
the surface and at the base of the convection zone, and $\Omega_{\rm
  pole}=\onehalf[\mean\Omega(r_1,\theta_0)+\mean\Omega(r_1,\pi-\theta_0)]$
is the average rotation rate between the latitudinal boundaries
on the outer radius.
In the Sun, $\Delta_\Omega^{(r)}>0$ and $\Delta_\Omega^{(\theta)}>0$,
which is what is
required for a solution to be classified as `solar-like'.

For high values of $\delta n$, i.e.\ for low radiative flux (Runs~A--C), 
the rotation profile is anti-solar with a negative radial
gradient of $\mean\Omega$ at all latitudes. The difference
$\Delta\mean\Omega$ between the equator and latitudes $\pm75\degr$ is
larger than $1.5\Omega_\odot$ in both cases. In Run~D the rotation
profile flips to solar-like.
Thus, the transition from the anti-solar to solar-like regime occurs
when $1.85 < \delta n < 2.0$, corresponding to $1.33 < \Co < 1.50$
in this set of runs. 
This is compatible with the results of \cite{GYMRW14}, who
found the transition at a {\em local} Rossby number of
around ${\rm Ro}_l\approx 1$ where 
${\rm Ro}_l\approx 2/\Co$.
As noted by
\cite{GYMRW14}, the transition is abrupt and occurs in a narrow
parameter range.
In Runs~D and E, with the highest Coriolis numbers, the rotation
profile is clearly solar-like. However, there are some interesting
features in Run E: a strong polar jet appears on the northern
hemisphere, and a decrease in $\mean\Omega$ is seen near the equator.
Such polar vortices have frequently been found in similar
simulations, see, e.g., \cite{KKBMT10,KMGBC11,KMB11}, and were
also found in rapidly rotating convection in relatively thin shells
\cite[e.g.][]{EMT00,GW12}.
  In the current setup this
  tendency is weaker, but we still occasionally observe polar jets
  (Runs E, B9, and B10; see also the left panel of
  Fig.~\ref{fig:pOmB10}.

We note that Run~A is expected to be
closest to the Sun with the highest convective energy flux.
Furthermore, our choice of $\PraSGS=0.25$ leads to a fairly large
contribution of SGS-flux within the convection zone, which in the Sun
is transported by convection. These results imply that the convective
velocities in the Sun would be even higher, leading to lower $\Co$ and
conditions that are more suitable for anti-solar differential rotation. There
are, however, hints that the velocities in simulations might be
significantly higher than those in the Sun \citep[cf.][]{HDS12,MFRT12}.

\subsection{Meridional circulation}
Figure~\ref{fig:pMC} shows the meridional circulation from
representative Runs~A and D.
In the relatively slowly rotating anti-solar Run~A, the flow is
concentrated in a single anti-clockwise cell mostly outside the
tangent cylinder with a peak amplitude of $90$\,m\,s$^{-1}$, which is
clearly higher than what is observed in the Sun
\citep[e.g.][]{ZK04}. 
In Run~D the circulation pattern extends to
higher latitudes and consists of several cells at low latitudes. The
cell at high latitudes is likely an artefact of the closed
$\theta$-boundary. 
A similar transition from multiple to single cells has been
observed before in different settings \citep[e.g.][]{KMB11,MCBB11,GWA13}.
The flow amplitude near the surface in Run~D is of the order
of $30\m\s^{-1}$, which is still somewhat higher than 
the $20\m\s^{-1}$ obtained from helioseismology
\citep{ZK04}. 

A single-cell poleward circulation with solar-like
rotation has been reported from simulations in spherical shells
with the ASH code by imposing a latitudinal entropy variation
on the bottom boundary \citep{Mie07,MBBBT11}.
In our spherical-wedge simulations, such a
circulation pattern in combination with a solar-like differential
rotation profile has so far occurred only as a transitory phenomenon
in runs that have not yet fully relaxed, and they typically end up in the
anti-solar regime.
Recent helioseismic studies suggest that the solar
meridional circulation pattern consists of several cells in radius and
possibly also in latitude \citep{ZBKDH13,STR13,KSJ14}, which is
also realized in our more rapidly rotating cases but is at odds
  with mean-field models of solar rotation \citep[e.g.][]{Re05,KR05}.

\subsection{Flow bistability}
We confirm recent results of \cite{GYMRW14}
that near the transition from solar-like to anti-solar
differential rotation, two stable solutions for the large-scale flow
exist for the same parameter values, only depending on the initial
conditions.

Our results for $\Delta_\Omega^{(r)}$ and $\Delta_\Omega^{(\theta)}$ are shown
in Fig.~\ref{fig:pDR} for three sets of models (cf.\ Table~\ref{tab:runs}). Firstly, we run models
from the initial conditions described in
Sect.~\ref{sec:initcond}. Furthermore, we run two additional sets
where we take a snapshot from an anti-solar and a
solar-like solution as initial conditions. In the last two sets we
vary the Rayleigh and Prandtl numbers by changing $\delta n$
and hence the radiative
conductivity $K(r)$, while keeping the other control parameters at fixed
values.
We find that for these choices of parameters and initial conditions, it
is more difficult to switch from anti-solar to solar-like differential
rotation than vice versa.
This is seen by comparing the $\delta n$ required for solar-like
solutions in the different sets of runs: in Set~D where we approach
from the rapid rotation regime, the switch occurs between $0.68 <
\Roc < 0.70$ ($1.27 < \Co < 1.37$). In the opposite case of Set~B,
where we approach from the anti-solar branch, the switch occurs
between $0.53 < \Roc < 0.55$ ($2.43 < \Co < 2.01$). In the case of
Runs~A--E that were run from scratch, we found $0.63 < \Roc < 0.65$
($1.33 < \Co < 1.50$). Thus, in terms of the Coriolis number the
bistability region extends farther into the anti-solar regime than the
solar-like one.
Physically, this might be related to the fact that in this case
the strength of the differential rotation is much larger (see
the two panels of Fig.~\ref{fig:pDR}).
We have considered a single value of the Taylor number in our
  study. We note that according to \cite{GYMRW14}, the size of the
  bistable region is wider with higher $\Ta$.

\begin{figure}[t]
\centering
\includegraphics[width=0.5\columnwidth]{pOm_C.eps}\includegraphics[width=0.5\columnwidth]{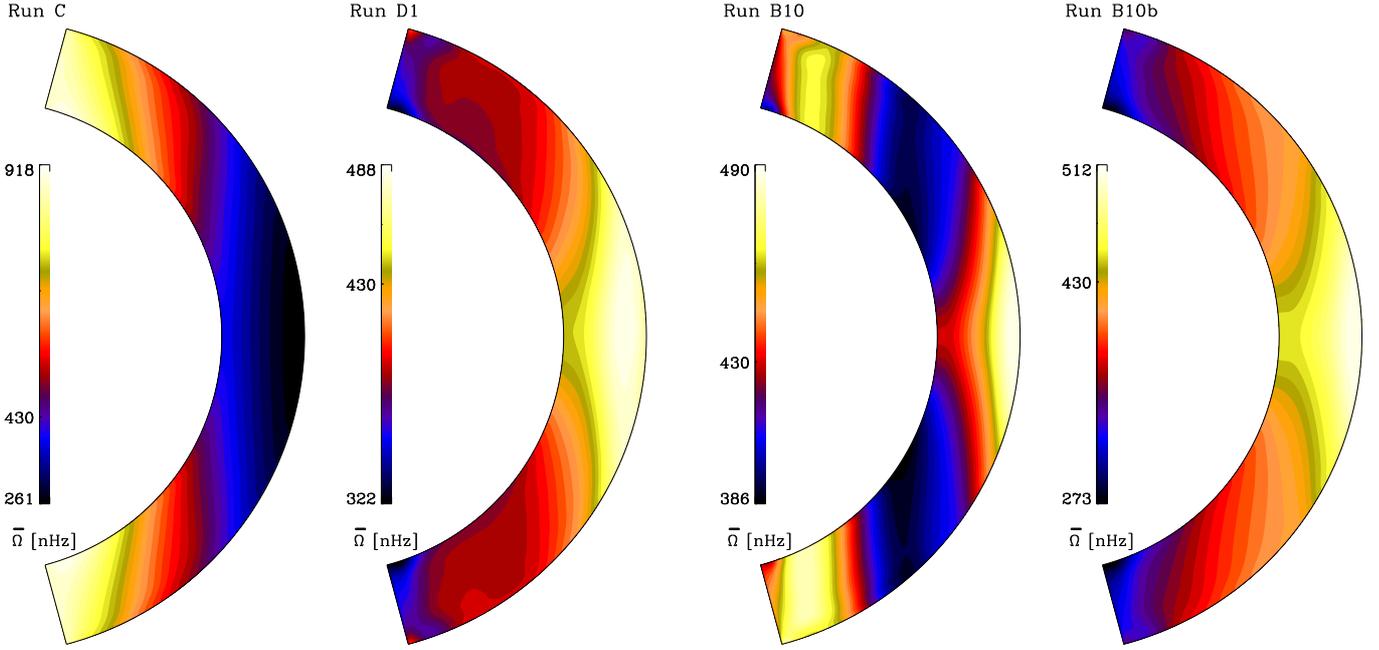}
\caption{Time-averaged rotation profiles from Runs~C and D1.}
\label{fig:pOmC}
\end{figure}

\begin{figure}[t]
\centering
\includegraphics[width=\columnwidth]{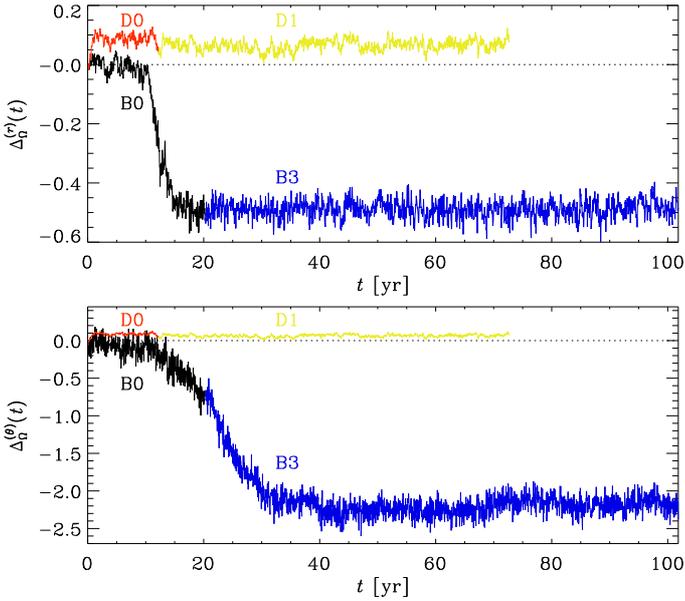}
\caption{Radial and latitudinal differential rotation as functions of
  time from Runs~D1 and B3 with the same parameters.}
\label{fig:pDRt}
\end{figure}

\begin{figure}[t]
\centering
\includegraphics[width=0.5\columnwidth]{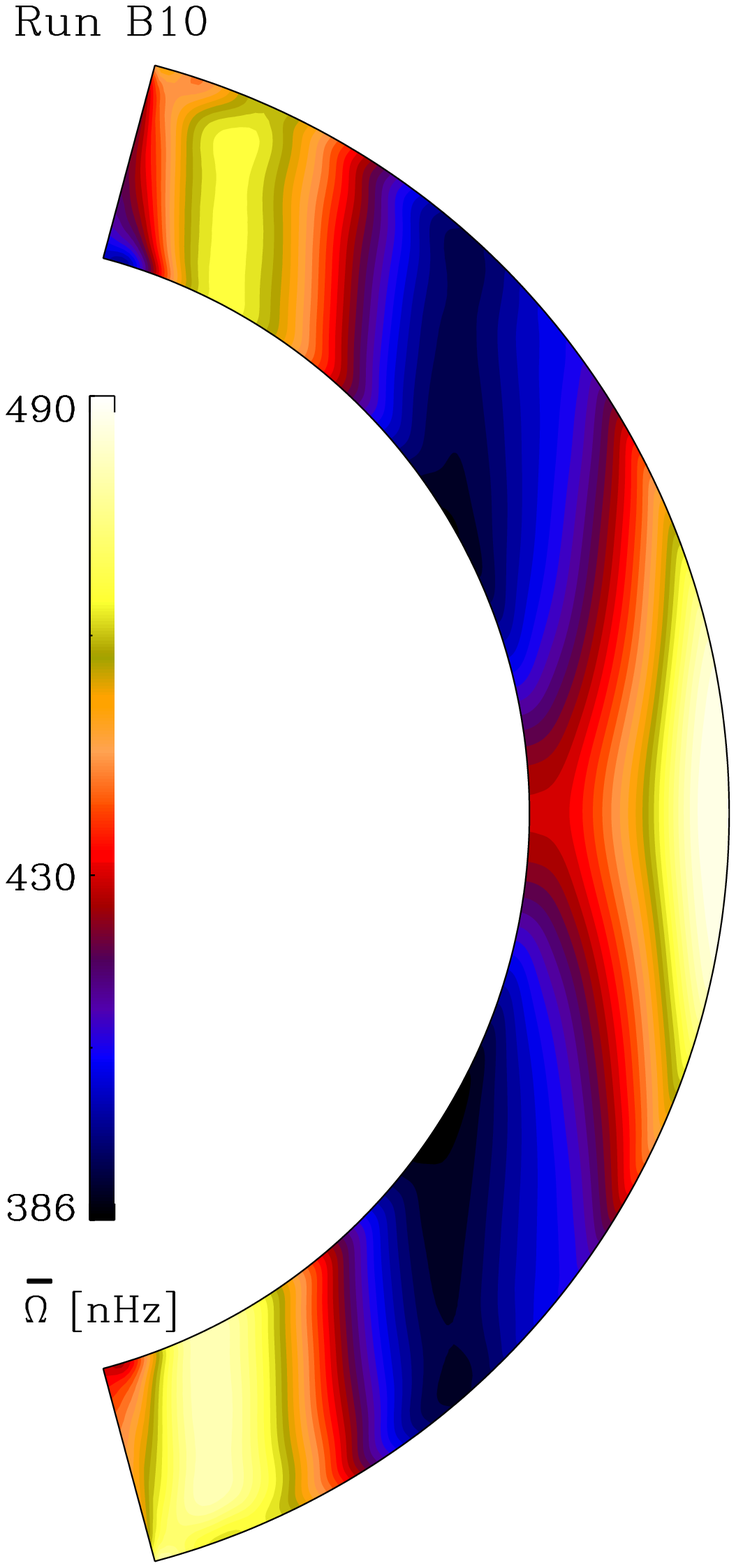}\includegraphics[width=0.5\columnwidth]{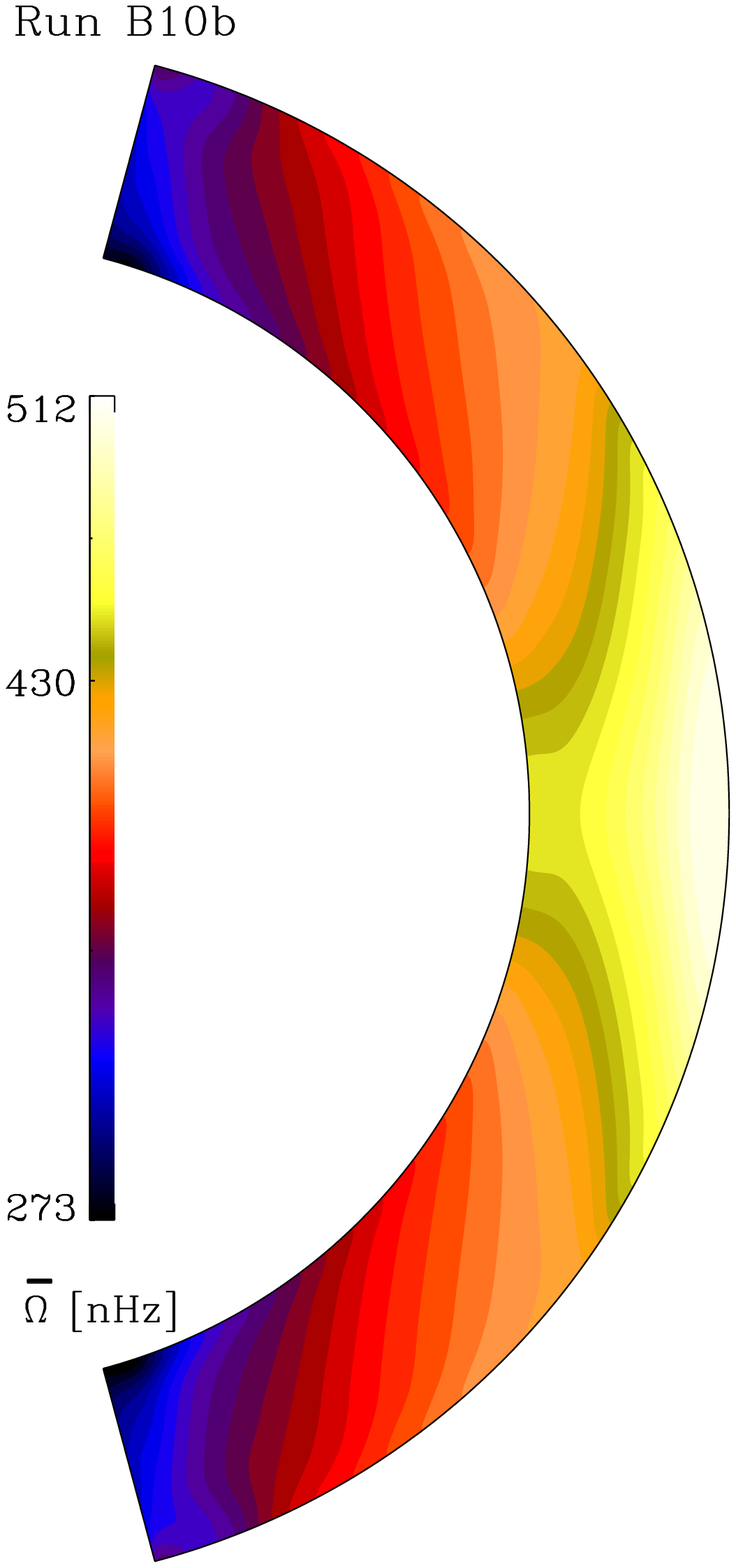}
\caption{Time-averaged rotation profiles from Runs~B10 and B10b.}
\label{fig:pOmB10}
\end{figure}

Figure~\ref{fig:pOmC} shows the rotation profiles from Runs~C and
D1 that have the same control parameters but different histories:
Run~C was run from the initial conditions described in
Section~\ref{sec:initcond}, whereas in Run~D1 we used the final
thermally relaxed state
of Run~D (=D0) as initial condition.
The resulting evolution of $\Delta_\Omega^{(r)}$ and
$\Delta_\Omega^{(\theta)}$ is shown in Fig.~\ref{fig:pDRt},
where we also show the corresponding results for Run~B3 with the
same input parameters as Run~D1 after restarting from Run~B (=B0).
Our simulations were typically run for roughly 100 years solar
time. By comparison, the viscous and SGS diffusion times,
$\tau_\nu=(\Delta r)^2/\nu$ and $\tau_{\rm SGS}=\tau_\nu\PraSGS$,
in our simulations are $10.5$ and $2.6$ years, respectively.

Rapidly rotating toroidal jets at high latitudes appear in many numerical
simulations of global scale convection
\citep[e.g.][]{METCGG00,KMB11,KMGBC11}. 
In the Sun the latitudinal gradient of $\mean\Omega$ is known to
  be monotonic in each hemisphere. Models based on solar differential
  rotation have also been adopted in stellar studies where
low latitude bands or polar vortices are ignored. Non-magnetic
mean-field models also tend to produce this type of
solution. However, the gas giants in the solar system have
alternating bands of slower and faster rotation, although in that case
it is not clear whether the convective layer is deep
\citep[e.g.][]{Busse76,HA07} or shallow \citep{KSHAH13}.
There are also theoretical and observational studies suggesting
similar profiles for fully convective M-dwarfs and brown dwarfs
\citep[e.g.][]{BW10,CBSea14}.

In Figure~\ref{fig:pOmB10} we show the rotation profiles obtained in
Runs~B10 and B10b with the same control parameters. Run~B10 was run
from a snapshot of Run~B0 exhibiting anti-solar differential
rotation.
This run now exhibits polar jets.
However, it has been suggested that these jets can be unstable
\citep{WJZ02}.
To investigate this possibility further, we
take a snapshot from B10 as initial for Run~B10b
and apply a relaxation term at high latitudes for the
azimuthally averaged $u_\phi$ that yields a monotonic latitudinal
gradient of $\mean{\Omega}$. The relaxation timescale is roughly four
days and the term is switched on for two weeks in solar time in the
beginning of the simulation. We find that
the resulting profile with a more solar-like monotonic behaviour is
also stable---at least for 50 years, which corresponds to roughly 
five viscous diffusion times.

\subsection{$\Lambda$ effect and turbulent viscosity}

\begin{figure*}[t]
\centering
\includegraphics[width=\textwidth]{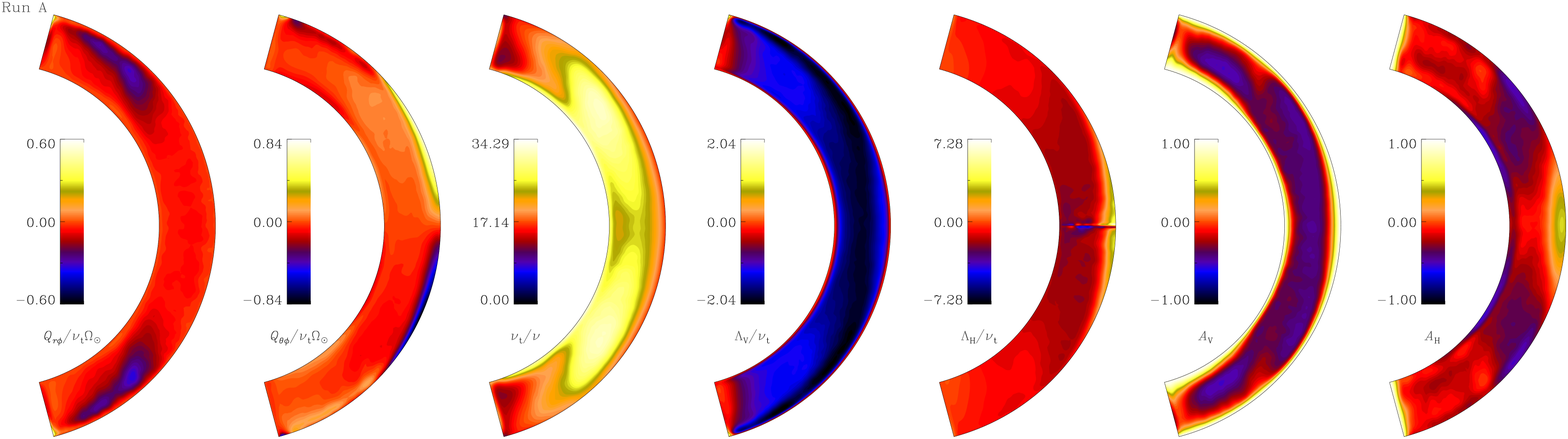}
\includegraphics[width=\textwidth]{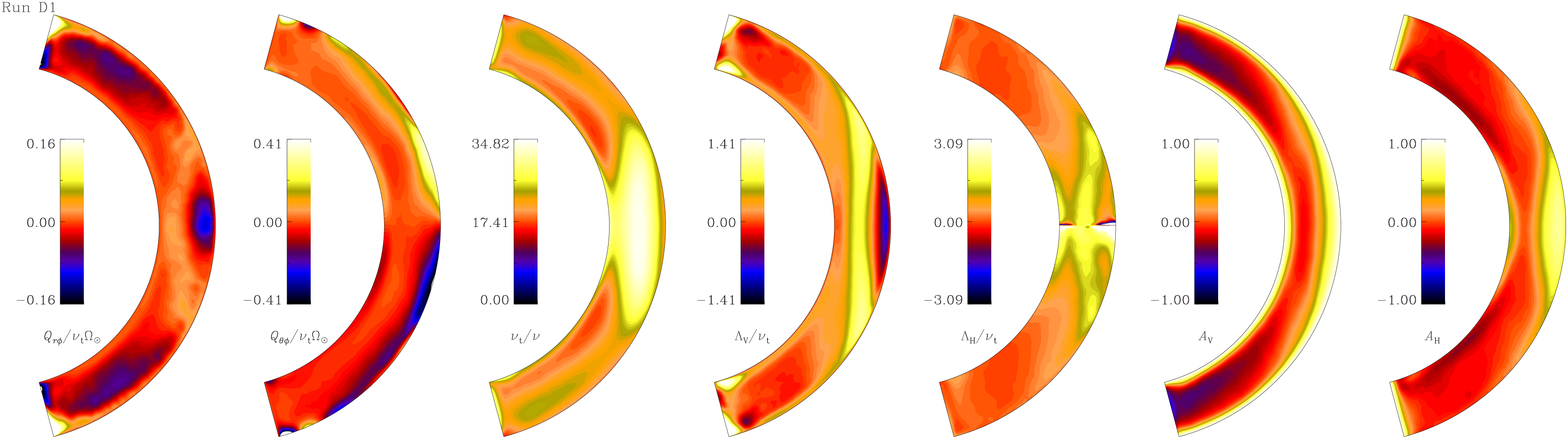}
\caption{From left to right: Time-averaged Reynolds stresses $\qrp$
  and $\qtp$ normalized by $\nut \Omega_\odot$, the turbulent viscosity
  divided by the molecular viscosity $\nut/\nu$, $\Lambda_{\rm V}$
  and $\Lambda_{\rm H}$ normalized by $\nut$, and the anisotropy 
  parameters $A_{\rm V}$ and $A_{\rm H}$. Top row: Run~A; bottom
  row: Run~D1. In the fifth column we only use data some degrees away
  from the equator so as to avoid the singularity associated with the
  division by $\cos\theta$. The contours in the lower row are
  oversaturated near the $\theta$-boundaries in order to highlight the
  features at lower latitudes.}
\label{fig:Reynolds}
\end{figure*}

The changes in the differential rotation should somehow be reflected in
similar changes in the underlying mechanism responsible for driving it,
which is the $\Lambda$ effect \citep{R80,R89}.
The $\Lambda$ effect corresponds to a rank three tensor that parameterizes
the non-diffusive contributions to the Reynolds stress, in addition to the
diffusive contributions that result from turbulent viscosity.
The Reynolds stress is given by $\qij=\mean{u_i' u_j'}$, where
$\bm{u}' = \bm{u} - \meanv{u}$ is the fluctuating velocity.
The relevant off-diagonal components can be written as
\begin{eqnarray}
\qrp &=& \LamV \sin\theta\mean\Omega - \nut r \sin \theta \frac{\pd \mean\Omega}{\pd r}, \label{equ:Rrp} \\
\qtp &=& \LamH \cos\theta\mean\Omega - \nut \sin \theta \frac{\pd \mean\Omega}{\pd \theta}, \label{equ:Rtp}
\end{eqnarray}
where $\LamV$ and $\LamH$ are the vertical and horizontal components
of the $\Lambda$ effect and $\nut$ is the turbulent
viscosity. Obviously, we cannot extract both effects self-consistently
from a single Reynolds stress component. Instead, we use a simple
mixing length formula to estimate the turbulent viscosity
\begin{equation}
\nut=\onethird \urms \alpha_{\rm MLT} H_p,
\end{equation}
where $\urms=\urms(r,\theta)$ varies across the meridional plane,
$\alpha_{\rm MLT}=1.7$ is taken for the mixing length parameter, and
$H_p=-(\pd \ln p/\pd r)^{-1}$ is the
pressure scale height. Similar methods have been applied in earlier
studies of
isotropically forced turbulence \citep{SKKL09} and convection
\citep{PTBNS93,KBKSN10} in Cartesian geometry, where anisotropy
is self-consistently produced by rotation and/or stratification.
In Fig.~\ref{fig:Reynolds} we present the results for Runs~A and D,
which are representative of anti-solar and solar-like rotation regimes.

In both cases,
the Reynolds stress responsible for radial transport of angular
momentum, $\qrp$, is negative at high latitudes. Somewhat surprisingly,
$\qrp$ is very small near the equator in the relatively slowly
rotating Run~A, but consistent with early results of \cite{RBMD94}.
In Run~D1, in the solar-like regime, a positive
contribution to the stress appears at low latitudes. The latter is
consistent with Cartesian simulations where the corresponding stress
changes sign at low latitudes in the rapid rotation regime
\citep{KKT04}. The horizontal stress, $\qtp$,
leads to angular momentum transport that is directed toward the
equator in both runs, except at high latitudes in Run~A where the
transport is toward the poles.

The mixing length estimate of the turbulent viscosity shows a decrease
at high latitudes in comparison to the equatorial regions in Run~A. In
Run~D1 there is a minimum at mid-latitudes. In both cases the maximum
value of $\nut/\nu$ is of the order of 35, which is reasonable given that the
Reynolds numbers in both runs are similar. We have here
neglected anisotropies that are caused by gravity and rotation that
play the roles of preferred directions in the system.
Thus, we do not expect the
details of the turbulent viscosity to be captured accurately. However,
the order of magnitude of $\nut/\nu\approx \Rey$ seems reasonable, so
we proceed to using this estimate to extract the $\Lambda$ effect.

Solving Eqs.~(\ref{equ:Rrp}) and (\ref{equ:Rtp}) for $\LamV$ and $\LamH$ yields
\begin{eqnarray}
\LamV &=& \frac{\qrp}{\sin\theta \; \mean\Omega} + \nut r \frac{\pd \ln \mean\Omega}{\pd r}, \label{equ:LamV} \\
\LamH &=& \frac{\qtp}{\cos\theta \; \mean\Omega} + \nut \tan\theta \frac{\pd \ln \mean\Omega}{\pd \theta}. \label{equ:LamH}
\end{eqnarray}
The profiles of $\LamV$ show some distinct differences between
both runs. It is mostly negative for the anti-solar Run~A and mostly
positive for the solar-like Run~D1.
Interestingly, these differences would not have been so obvious
if we had directly compared the vertical stresses $\qrp$ of both runs.
This highlights the usefulness of employing Eq.~(\ref{equ:LamV}),
even though this involves the uncertainty of estimating $\nut$.
Likewise, while the horizontal stresses $\qtp$ for Runs~A and D1
show some similarities,
the profiles of $\LamH$ also show differences between both runs, and
it is mostly negative for the anti-solar Runs~A and mostly
positive for the solar-like Run~D1.

According to mean-field theory \citep{R80}, coefficients
$\LamV$ and $\LamH$ are related to the anisotropy parameters 
\begin{eqnarray}
A_{\rm V} = \frac{Q_{\phi\phi} - Q_{rr}}{Q_{\phi\phi} + Q_{rr}},\ \ \ A_{\rm H} = \frac{Q_{\phi\phi} - Q_{\theta\theta}}{Q_{\phi\phi} + Q_{\theta\theta}},
\end{eqnarray}
via $\Lambda_{\rm V} \approx 2 \tau A_{\rm V}$ and
$\Lambda_{\rm H} \approx 2 \tau A_{\rm H}$, where $\tau$ is the
correlation time of the turbulence.
Profiles of $A_{\rm V}$ and $A_{\rm H}$ are shown in the
last two columns of Fig.~(\ref{fig:Reynolds}).
In Run~A, both $\Lambda_{\rm V}$ and $A_{\rm V}$ are mostly
negative, while both are mostly positive for Run~D1, in broad
agreement with the {\em radial} differential rotation gradient.
Also $\Lambda_{\rm H}$ and $A_{\rm H}$ are mostly negative in Run~A
and mostly positive in Run~D1, again in broad agreement with the
{\em latitudinal} differential rotation.

According to first-order
smoothing results \citep[e.g.][]{KR95,KR05}, $A_{\rm V}$ is always
negative due to the simplified turbulence model used. This,
however, is not always the case in simulations \citep[e.g.][]{KKT04}.
Indeed, $A_{\rm V}$ is mostly negative in Run~A and attains
positive values near the equator in Run~D1, in qualitative accordance
to our estimates of $\Lambda_{\rm V}$ for each case. Similarly we find
that the sign of $A_{\rm H}$ agrees mostly with that of $\Lambda_{\rm
  H}$. It is remarkable that the results for the $\Lambda$ effect are
consistent with those of the anisotropy parameters given our use of a
very simple approximation for $\nut$. Our results for $A_{\rm V}$ and
$A_{\rm H}$ also indicate that the differential rotation has a
significant impact on the properties of turbulence.

For the run with solar-like rotation (Run~D), $\LamH$ is concentrated near
the equator and close to the upper boundary. A similar concentration
near the equator has also been observed in earlier studies
\citep[e.g.][]{Chan01,KKT04} and can be partly explained by the banana
cells near the equator \citep{KMGBC11}.

\section{Conclusions}

Simulations of mildly turbulent three-dimensional convection in
spherical wedges have allowed us to study aspects of differential rotation
relevant to the Sun.
In our most solar-like model, Run~A, we found anti-solar
differential rotation. Decreasing the convective velocities by
increasing the radiative conductivity, thereby also increasing the
Coriolis number, we found a threshold after which the rotation changes
to a solar-like profile.
In agreement with the recent finding of \cite{GYMRW14} using Boussinesq
convection, our stratified and compressible models near the threshold
between both states confirm the existence of bistability in that
the question of anti-solar
or solar-like rotation depends on the initial condition or the history
of the run.
We also confirmed earlier findings showing that
high-latitude toroidal jets are possible in runs
where solar-like solutions are observed with the same parameters.
This could be significant for stellar rotation profiles that are
naively assumed to have a monotonic latitudinal gradient of angular
velocity. These jets may well be asymmetric and present only at one
of the poles \citep[see also][]{HA07,JK09}.

Another interesting but speculative conjecture is
that the solar differential rotation is also the result of
bistability. Our Runs~A and B with energy fluxes closest to what we
would expect from the Sun i.e.\ where resolved convection transports
most of the total flux, show anti-solar differential rotation.
Making the model more realistic with respect to the Sun
requires that we (i) decrease the SGS-flux, which now contributes more than
ten per cent of the total, (ii) increase the density
stratification by a factor of five, and (iii) increase the Reynolds number.
All of these factors are likely to lead to higher
convective velocities and thus lower Coriolis number, in turn leading to even
more suitable conditions for anti-solar differential rotation. We
know, however, that the Sun rotated much more rapidly in its youth,
which favours a solar-like rotation profile. Furthermore, the results
of \cite{GYMRW14} suggest that the range of parameters in which
bistable solutions are possible increases as the Taylor number
increases. These two facts lend some credence to the conjecture that
the Sun is in a bistable regime. Clearly, further research into this
matter is required.

\begin{acknowledgements}
We thank the referee for a thorough report and constructive criticism.
The simulations were performed using the supercomputers hosted by the CSC
-- IT Center for Science Ltd.\ in Espoo, Finland, which is administered
by the Finnish Ministry of Education. Financial support from the
Academy of Finland grants No.\ 136189, 140970, 272786 (PJK)
and 272157 to the ReSoLVE Centre of Excellence (MJM),
the University of Helsinki research project `Active Suns',
as well as the Swedish Research Council grants 621-2011-5076 and 2012-5797
and the European Research Council under the AstroDyn Research Project
227952, are acknowledged, as well as the HPC-Europa2 project, funded by
the European Commission - DG Research in the Seventh Framework
Programme under grant agreement No.\ 228398.
\end{acknowledgements}

\bibliographystyle{aa}
\bibliography{paper}


\end{document}